\documentclass[preprint,amsfont,amssymb,amsmath, showpacs,balancelastpage, nofootinbib]{revtex4-2}

\usepackage{amsthm}

\newcommand{\tl}{\triangleleft}

\usepackage[utf8]{inputenc}
\usepackage{amsmath,amssymb,amsfonts,stmaryrd}
\usepackage{physics}
\usepackage{dsfont}
\usepackage{graphicx}
\usepackage{xcolor}
\usepackage{graphicx}
\usepackage{cancel}
\usepackage{natbib}
\usepackage{color}
\usepackage{tikz}
\usetikzlibrary{matrix}
\usepackage{mathrsfs}
\usepackage{relsize}
\usepackage{multirow}
\usepackage[linktocpage]{hyperref}
\usepackage[all]{xy}
\hypersetup{colorlinks=true,citecolor=blue,linkcolor=blue, urlcolor=blue, breaklinks=true}
\usepackage[scr=boondox]{mathalpha}

\usepackage{comment}

\usepackage{bm} 
\usepackage{subcaption} 
\usepackage{environ}
\usepackage{url}
\usepackage[margin=0.75in]{geometry}

\usepackage{mathtools}

\newcommand{\Z}{{\mathbb Z}}

\def\U{\mathrm{U}}

\newcommand{\pt}{\text{pt}}

\newcommand{\SO}{\text{SO}}
\newcommand{\SU}{\text{SU}}

\newcommand{\Tt}{\mathrm{t}}

\newcommand{\HH}{\mathcal{H}}

\newcommand{\eqnref}[1]{Eq.~(\ref{#1})}

\NewEnviron{eqs}{%
\begin{equation}\begin{split}
    \BODY
\end{split}\end{equation}}

\theoremstyle{definition}

\theoremstyle{remark}

\begin{document}
\title{Anomalous continuous symmetries and quantum topology of Goldstone modes}
\author{Naren Manjunath}
\affiliation{Perimeter Institute for Theoretical Physics, Waterloo, Ontario N2L 2Y5, Canada}

\author{Dominic V. Else}
\affiliation{Perimeter Institute for Theoretical Physics, Waterloo, Ontario N2L 2Y5, Canada}

\begin{abstract}
We consider systems in which a continuous symmetry $G$, which may be anomalous, is spontaneously broken to an anomaly-free subgroup $H$ such that the effective action for the Goldstone modes contains topologically non-trivial terms. If the original system has trivial $G$ anomaly, it is known that the possible topological terms are fully determined by SPT or SET invariants of the residual $H$ symmetry. Here we address the more general setting in which the $G$ symmetry has an anomaly. We argue that in general, the appropriate concept to consider is the ``compatibility relation'' between the Goldstone invariants and the $G$ anomaly. In the case where the Goldstone modes can be gapped out to obtain invertible families (i.e. without any topological order), we give an explicit mathematical scheme to construct the desired compatibility relation. We also address the case where gapping out the Goldstone modes leads to a family of topologically ordered states. We discuss several examples including the canonical Thouless pump, the quantum Hall ferromagnet, pumps arising from breaking $\U(1)$ symmetry at the boundary of topological insulators in two and three dimensions, and pumps classified by the higher Chern number.

\end{abstract}
\maketitle
\newpage
\tableofcontents

\section{Introduction}
In this paper we consider many-body quantum systems with a continuous symmetry $G$, which is spontaneously broken down to a subgroup $H$. The associated order parameter then lives in the quotient space $\Lambda := G/H$. The resulting system is gapless, due to the presence of Goldstone modes. The fact that there can be non-trivial `quantum topology' associated to these modes has long been appreciated in both quantum field theory and condensed matter contexts \cite{WessZumino1971,Abanov1999-NLSM,Brauner2014,Yonekura2020Anomaly,else2021goldstone}. Roughly speaking, this corresponds to the low-energy effective action for the Goldstone modes containing  quantized topological terms, reflecting the non-trivial quantum many-body entanglement carried by the gapped degrees of freedom. A consequence of these topological terms could be, for example, that smooth topological configurations of the order parameter (e.g. skyrmions) bind quantized charge \cite{Sondhi1993skyrmion,Moon1995,Zhang2019Moire,Khalaf2020ChargedSA,Chatterjee2020TBG,Wu2020moire} or that singular configurations such as vortices bind topologically protected gapless modes. Phases of matter in which the Goldstone modes carry such topological terms were dubbed `topological Goldstone phases' in Ref.~\cite{else2021goldstone}. 

In this paper, we will give the general framework to understand when such topological Goldstone phases of matter can occur. We build on Ref.~\cite{else2021goldstone}, which considered the case where $G$ is non-anomalous, and then showed that the topological invariant of the Goldstone modes can always be related to an SPT/SET invariant of the residual $H$ symmetry. Here we consider the more general case, in which $G$ can be anomalous; in other words, the system could exist on the boundary of a $G$-SPT in one higher dimension. We develop a unified framework to characterize the Goldstone modes in terms of the $G$ anomaly (if it is nontrivial), or in terms of a topological invariant of $H$, when there is no $G$ anomaly.

Specifically, as in Ref.~\cite{else2021goldstone}, we consider systems in which the \emph{only} gapless degrees of freedom are the Goldstone modes (thus, we do not consider, for example, spontaneous symmetry-breaking phases in metals, in which the Goldstone modes co-exist with gapless fermions). Therefore, if we add a weak perturbation to the Hamiltonian that explicitly breaks the $G$ symmetry down to $H$, its only effect will be to weakly gap out the Goldstone modes. One then obtains a family of gapped ground states parameterized by the order parameter space $G/H$ \cite{else2021goldstone}. This allows us to invoke the body of work on understanding the classification under homotopy equivalence of parameterized families of gapped quantum many-body systems \cite{Cordova2020anomcoupling,Hsin2020berry,Kapustin2020HBC,Kapustin2020Thouless,Wen2021HBC,Beaudry2023Parametrized,Debray2023SB}. The topological terms for the Goldstone modes correspond to topological invariants for the gapped family.

The question we address can then be formulated as follows. Let $\Lambda$ be some parameter space, and suppose a group $G$ acts transitively on $\Lambda$ (that is, any two elements of $\Lambda$ can be related by the action of some element of $G$). Then $\Lambda$ is equivalent to a quotient space $G/H$ for some subgroup $H$. Now suppose that we have some local (anti-)unitary representation $U(g)$ of $G$ on a quantum many-body system. This representation could be on-site, or it could be non-on-site, in which case it could potentially have an anomaly. Now consider a family of gapped states $\ket{\Psi(\lambda)}$ parameterized by $\Lambda$ such that
\begin{equation}
U(g) \ket{\Psi(\lambda)} = \ket{\Psi(g\lambda)}.
\end{equation}
What is the relation between the $G$-anomaly of the local unitary representation and the topological invariants of the family? The answers to this question will also apply more generally, beyond the specific application of spontaneous symmetry breaking. However, in this paper we will mainly focus on the spontaneous symmtery breaking application.

Our results can be viewed as somewhat similar in spirit to Ref.~\cite{Debray2023SB}. However, unlike Ref.~\cite{Debray2023SB}, we do not restrict ourselves to Lorentz-invariant systems, and we do not require the parameter space $\Lambda$ to be a sphere. Meanwhile, unlike this work, Ref.~\cite{Debray2023SB} did not require the action of $G$ on $\Lambda$ to be transitive. These differences will result in us obtaining quite a different structure from Ref.~\cite{Debray2023SB}.

 The outline of this paper is as follows. In Section~\ref{sec:Summary} we present our main result, which is that there exists a `compatibility relation' between a given $G$ anomaly and a family obtained by spontaneously breaking $G$, then gapping out the Goldstone modes. We will discuss both invertible and topologically ordered families. In Sec.~\ref{sec:Apps} we discuss a number of applications illustrating this compatibility relation in condensed matter systems. In Secs.~\ref{sec:GC-Classif} and \ref{sec:MainTechRes} we review the generalized cohomology classification for invertible states and invertible families and explain how to mathematically derive the compatibility relation using these classifications. We give a further discussion of non-invertible families in Sec.~\ref{sec:MainRes-noninv}, and then conclude in Sec.~\ref{sec:Disc}.

\section{Summary of results}\label{sec:Summary}

\subsection{Set-up}\label{subsec:setup}
We consider a symmetry $G$ that is spontaneously broken to a subgroup $H$. Thus, the order parameter takes values in the space $\Lambda := G/H$. We can consider the family of ground states parameterized by $\Lambda$. 
Let the symmetry preserved by the whole family be $H_0$, which must be a normal subgroup of $H$.\footnote{More specifically, $H_0$ is the largest normal subgroup of $H$ which is also a normal subgroup of $G$.}
We write $\{\text{$H_0$-families over $\Lambda$}\}_d$ for the classification of families over a space $\Lambda$ with $H_0$ symmetry.

We can distinguish between \emph{invertible} and \emph{non-invertible} families. First, a gapped ground state $\ket{\Psi}$ is said to be invertible if there exists another gapped ground state $\ket{\Psi^{-1}}$ such that $\ket{\Psi} \otimes \ket{\Psi^{-1}}$ can be adiabatically connected to a product state. This also implies that the ground state is non-degenerate when defined on any closed manifold. We say a gapped ground state is non-invertible (or topologically ordered) if it is not invertible.  We say that a family is invertible (non-invertible) if the individual ground states within a family are invertible (non-invertible) states. 
If we focus on invertible famlies, then $\{\text{Invertible $H_0$-families over $\Lambda$}\}_d$ is an Abelian group under stacking, and one can argue that there is a direct product decomposition
\begin{equation}
\label{eq:family_product}
\{\text{Invertible $H_0$-families over $\Lambda$}\}_d = \{\text{$H_0$-inv} \}_d \times \{ \text{$H_0$-pumps over $\Lambda$} \}_d
\end{equation}
Here \{$H_0$-inv\} is the classification of invertible ground states in $d$ dimensions with $H_0$ symmetry. (A notational remark: we will use the terms `invertible states with $H_0$ symmetry', `$H_0$-invertible states', and `$H_0$-SPTs' interchangeably in this paper, and in every case we include the states that remain nontrivial when we forget the $H_0$ symmetry.) In the context of the decomposition \eqnref{eq:family_product}, these correspond to \emph{constant} families in which the ground state is the same for each value of the parameter space $\Lambda$. Meanwhile, the $H_0$-pumps over $\Lambda$ are defined to be the families in which the ground state for any fixed value of the parameter is in the trivial $H_0$-SPT.

The motivation for the ``pump'' terminology comes from the observation that when $\Lambda = S^1$, we can vary the order parameter in a time-dependent way corresponding to a non-trivial winding around the circle as a function of time, and the associated family invariant generally corresponds to the pumping of an $H_0$-symmetric invertible state to the boundary during this process. Families over higher dimensional manifolds can often be interpreted as ``higher-order pumps''. While not all family invariants can necessarily be interpreted this way, we will use the term ``pump'' in general.

We are interested in the relationship between $\{\text{$H_0$-families over $\Lambda$}\}_d$
and two other classifications. Firstly, the $G$-symmetry may act in an anomalous way on the system. Thus, we define $\{\text{$G$-inv}\}_{d+1}$ to be the Abelian group that classifies the anomalies of a $G$ symmetry in $d$ spatial dimensions (the notation reflects the fact that this is equivalent to the classification of invertible phases with $G$ symmetry in $d+1$ spatial dimensions).

Secondly, if we fix the order parameter to a particular value, then we have a ground state with a symmetry $H$. Thus, we can ask where this ground state falls in $ \{  \text{$H$-states} \}_d$, which is the classification of $H$-symmetric ground states in $d$ spatial dimensions. If we focus only on the cases where the ground states are invertible, then we can instead consider $ \{  \text{$H$-inv} \}_d$, the Abelian group which classifies $H$-symmetric invertible states in $d$ spatial dimensions.

\subsection{Review: the case of no $G$-anomaly}
The main result of Ref.~\cite{else2021goldstone} was as follows. Suppose that the symmetry $G$ has an on-site action (i.e.\ there is no $G$-anomaly). Let $\ket{\Psi(\lambda)}$ be the family of ground states parameterized by the order parameter $\lambda \in \Lambda$. If we fix some particular $\lambda_* \in \Lambda$, which is invariant under the subgroup $H \leq G$, then we can assign the state $\ket{\Psi(\lambda_*)}$ to an element $s \in \{ \text{$H$-states} \}_d$. The result is that the $H_0$-family invariant of $\ket{\Psi(\lambda)}$ is \emph{completely} determined by $s$.

 A simple argument for this result is as follows. Let us fix the action $U(g)$ of $G$ on the Hilbert space of the system. The family $\ket{\Psi(\lambda)}$ is $G$-equivariant, in the sense that $U(g) \ket{\Psi(\Lambda)} = \ket{\Psi(g\Lambda)}$ (up to global phases). Given that the $G$-action on $\Lambda$ is transitive (i.e.\ any two elements of $\Lambda$ are related to each other by the action of some group element), it follows that we can reconstruct the whole family $\ket{\Psi(\lambda)}$ from $\ket{\Psi(\lambda_*)}$. It follows that each $H$-invariant gapped state determines an $H_0$-invariant family. The one thing that this argument does not show is that the map from $\{ \text{$H$-states} \}_d \to \{ \text{$H_0$-families} \}_d$ is independent of the choice of representation $U(g)$, provided that it is on-site; however, this is certainly plausible and follows from the general arguments of Ref.~\cite{else2021goldstone}. Also, note that this argument does not actually depend on $U(g)$ being on-site; however, in the case of an anomalous action of the symmetry, the family that one can obtains will in general depend on the anomaly.

A simple example of these general considerations in the non-anomalous case is that of a zero-dimensional spin in a magnetic field $\vec{B}$, which at fixed $|\vec{B}| >0$ is a $\Lambda = S^2$-valued parameter. At $|\vec{B}|=0$, the system has a $G = \SO(3)$ spin rotation symmetry, which we can imagine is spontaneously broken down to $H = \SO(2)$ by any nonzero value of the field.  The symmetry $H_0$ is trivial, since each choice of $\vec{B}$ breaks $G$ down to a different $\SO(2)$ subgroup. [Although spontaneous symmetry breaking is not meaningful for an isolated system in $d=0$, we can consider a zero-dimensional quantum dot which is in proximity to a higher-dimensional substrate. Then it is meaningful to talk of the symmetry of the substrate (and by the proximity effect, the symmetry of the dot) being spontaneously broken.]

In this case, the family invariant is given by the Chern number of the Berry connection associated to the ground state by varying $\vec{B}$: it equals $2 s \in \Z$ where $s$ is the spin, and classifies the different families. Now, a spin with half-integer $s$ carries $G$-anomaly, which in $d=0$ just means it transforms projectively under $\SO(3)$. We will discuss this case further below. On the other hand, an integer $s$ spin is non-anomalous. In this case, the ground state of an integer $s$ system with magnetic field $\vec{B}$ along $\hat{z}$ has $S_z$ eigenvalue $m_s = s$, where $S_z$ generates an $\SO(2)$ subgroup identified with $H$. Thus the state carries integer charge under $\SO(2)$, which defines an $H$-SPT in $d=0$. Notice also that not all families over $S^2$ can be obtained from symmetry-breaking in the case where the $\SO(3)$ symmetry is non-anomalous, because one only obtains even integer Chern numbers in this case.

\subsection{A simple anomalous case}
\label{subsec:simple_anomaly}
Let us first consider the simple case of a direct product symmetry group $G = \hat{G} \times H$ which is broken down to $H$. In this case we have $H_0 = H$ and $\Lambda = \hat{G}$. In the results of Ref.~\cite{else2021goldstone} one can show that for this case, if $G$ is non-anomalous there will never be any non-trivial topological invariants carried by the Goldstone modes.
More precisely, the $H$-symmetric family over $\Lambda$ can be continuously deformed to a constant family.

More generally, we will show in this paper that in this simple case, if one does require the $G$ symmetry to be anomalous, then there is a group homomorphism
\begin{equation}\label{eq:simple_anomaly}
    \varphi : \text{\{$G$-inv\}}_{d+1}^H \to \text{ \{$H$-pumps over $\Lambda$ \}}_d
\end{equation}
where $\text{\{$G$-inv\}}_{d+1}^H$ denotes the subgroup of $\text{\{$G$-inv\}}_{d+1}$ corresponding to the anomalies that become trivial upon restricting to $H$.

The implications of this map are as follows. Let $\ket{\Psi(\lambda)}$, $\lambda \in \Lambda$ be the family of ground states in $d$ spatial dimensions obtained by spontaneously breaking $G$ to $H$ in the presence of $G$-anomaly $\alpha \in \{ \text{$G$-inv} \}_{d+1}^H$. Let $\varphi(\alpha) \in \{ \text{$H$-pumps over $\Lambda$} \}$ be the image of the anomaly under $\varphi$. Now introduce some auxiliary system, also in $d$ spatial dimensions, on which $H$ acts on-site (i.e.\ non-anomalously) and $\hat{G}$ acts trivially, and let $| \Phi_{\varphi(\alpha)} (\lambda) \rangle$ be a family of states in the auxiliary system which realizes the pump class $\varphi(\alpha)$. Then, if we introduce a state $\ket{0}$ which is a trivial $H$-invariant product state in the auxiliary system, what we find is that $\ket{\Psi(\lambda)} \otimes \ket{0}$ can be continuously deformed to
\begin{equation}
    \ket{\Psi(\lambda_*)} \otimes \ket{\Phi_{\varphi(\alpha)}(\lambda)}.
\end{equation}
In other words, the family $\ket{\Psi(\lambda)}$ is equivalent to the stack of a constant family with an $H$-pump determined by the $G$-anomaly.

These results apply whether $\ket{\Psi(\lambda_*)}$ is invertible or non-invertible. However, it is worth noting that since many families of non-invertible states cannot be realized as the stack of a constant family with a pump, it follows that such families cannot be realized as topological Goldstone phases in the spontaneous symmetry breaking pattern of the current subsection. See Sec.~\ref{sec:AnyonPerm} for a concrete example involving anyon-permuting families.

\subsection{General invertible case}
\label{subsec:general_invertible}

The situation for the most general case is more complicated, because it needs to interpolate between the results from both Ref.~\cite{else2021goldstone}, in which the Goldstone invariants are determined by the SPT/SET of the residual symmetry $H$, and Section \ref{subsec:simple_anomaly}, in which the Goldstone invariants are determined by the anomaly. In this subsection we will focus on the case where the ground states are invertible. We discuss the mathematical classification of invertible families further in Sec.~\ref{sec:MainTechRes}.

Before we proceed, there is a subtlety that we need to mention (which in fact was already present in the case studied in Section \ref{subsec:simple_anomaly}, but we formulated the discussion there in such a way as to sidestep the issue). In the presence of a $G$-anomaly, in general there is not a canonical way to identify the $H_0$-SPT carried by a particular $H_0$-symmetric ground state. One way to explain this is as follows. We can think of a system with a $G$-anomaly in $d$ spatial dimensions as existing on the boundary of a $G$-SPT in $d+1$ spatial dimensions. If the $G$-SPT is to be non-trivial, then its wavefunction must not be a product state. On the other hand, in order to get a gapped ground state on the boundary when we break $G$ down to $H$, it must be the case that $H$, and in particular $H_0$, is anomaly-free, and therefore the $(d+1)$-dimensional SPT is in the trivial SPT with respect to $H_0$. In order to be able to identify a boundary state with a $d$-dimensional $H_0$-SPT, we need to apply an $H_0$-symmetric local unitary to disentangle the $(d+1)$-dimensional bulk. However, there is not a unique choice for such a local unitary, and in particular, different choices can differ by pumping a $d$-dimensional $H_0$-SPT to the boundary. For this reason, the different classes of $H_0$-symmetric invertible ground states on the boundary must be viewed as a torsor over $H_0$-SPTs.

Now, consider a system in $d$ dimensions with a particular (anomalous) realization of a $G$ symmetry. If we consider the families over $\Lambda$ that are invariant under the $H_0$ subgroup of $G$, then the torsoriality that we just mentioned implies that we cannot canonically identify the classification of such families with \eqnref{eq:family_product}. However, the problem lies only with the first factor. Therefore, for a given $H_0$-symmetric invertible family $f$ over $\Lambda$ in the $G$-anomalous system, we can canonically define $\pi(f) \in \{ \text{$H_0$-pumps over $\Lambda$}\}$. Roughly speaking, we are quotienting out by the first factor of \eqnref{eq:family_product}. 

Next we want to consider the relation between such pumps and the $G$-anomaly. However, unlike the case studied in Section \ref{subsec:simple_anomaly}, the relationship will not take the form of a direct mapping from $G$-anomalies into $H_0$-pumps over $\Lambda$. Indeed, even in the case where the $G$-anomaly is trivial, we know from the results of Ref.~\cite{else2021goldstone} that we can have different $H_0$-pumps depending on the $H$-SPT carried by the symmetry-breaking ground states. Instead, the relationship will take the form of a \emph{compatibility relation}. We make the following definition:

\begin{quote}
    Let $p \in \{ \text{$H_0$-pumps over $\Lambda$} \}_d$, and let $\alpha \in \{ \text{$G$-inv}\}_{d+1}$. We say that $p$ and $\alpha$ are \emph{compatible} if there exists a system in $d$-dimensions with $G$-anomaly $\alpha$ and a family $f$ realized by spontaneously breaking $G$ to $H$, such that $\pi(f) = p$.
\end{quote}
Symbolically, if $p$ and $\alpha$ are compatible, we will write ``$p \tl \alpha$''. In mathematical terminology, this defines what is known as a ``binary relation''. Note that this relation should be compatible with the group operation (which physically corresponds to stacking). Specifically, if we write the group operation additively, then we must have:
\begin{itemize}
    \item $0 \tl 0$ (where the two $0$'s correspond to the trivial pump and the trivial anomaly respectively).
    \item If $p \tl \alpha$ then $-p \tl -\alpha$.
    \item If $p \tl \alpha$ and $p' \tl \alpha'$ then $p + p' \tl \alpha + \alpha'$.
\end{itemize}

Returning to the example of a spin in a magnetic field, the anomalies are $\Z_2$ classified (with spin $s$ carrying the anomaly $\alpha = 2s$ mod 2), while the families are classified by their Chern number, which is $p_C = 2s \in \Z$. In this case, the compatibility relation is that $p_C \tl \alpha$ if and only if $p_C = \alpha \mod 2$: the trivial and non-trivial anomalies are compatible with even and odd Chern number, respectively. 

The remainder of the paper will be mainly be devoted to describing this compatibility relation. For now we will just mention one property that it satisfies: if $\alpha \in \{ \text{$G$-inv} \}_{d+1}^H$ (recall that this comprises the $G$-anomalies that are trivial upon restricting to $H$), there is at least one $p \in \{ \text{$H_0$-pumps over $\Lambda$} \}_d$ such that $p \tl \alpha$.  Indeed, this comes about because if $\alpha \in \{ \text{$G$-inv} \}_{d+1}^H$, then there should be no obstruction to having a trivial ground state upon explicitly breaking $G$ down to $H$, and we can then reconstruct a $G$-equivariant family over $\Lambda$ just by acting on this state with $G$ symmetry.

We can derive the results of Section \ref{subsec:simple_anomaly} as a corollary. Indeed, in the case where $G = \hat{G} \times H$ as considered there, there are no non-trivial $H_0$-pumps compatible with the trivial anomaly, as a consequence of the results of Ref.~\cite{else2021goldstone}. Therefore, for any $\alpha \in \{ \text{$G$-inv} \}_{d+1}$, there is a unique $p \in \{ \text{$H_0$-pumps} \}_d$ such that $p \tl \alpha$, because if there are $p,p' \in \{ \text{$H_0$-pumps} \}_d$ such that $p \tl \alpha$ and $p' \tl \alpha$, then $p - p' \tl 0$ which implies $p = p'$. Therefore, we define the homomorphism $\varphi$ in \eqnref{eq:simple_anomaly} such that $\varphi(\alpha)$ is the unique $p \in \{ \text{$H_0$-pumps} \}_d$ such that $p \tl \alpha$. (The fact that it is a homomorphism follows from the properties of $\tl$ stated above).

Finally, let us remark that while the torsoriality issues described above prevent us from meaningfully lifting the compatibility relation on pumps to a compatibility relation on families, we can still make statements that involve comparing two families. Suppose we fix a particular anomalous $G$-action on the system, and then consider two $H_0$-symmetric families $f,f'$ both obtained by spontaneously breaking $G$ to $H$. Then one can argue that $f'$ is necessarily deformable to a stack $f \otimes f_0$, where $f_0$ is an $H_0$-symmetric invertible family obtained by spontaneously breaking $G$ to $H$ in a system with an \emph{on-site} action of $G$. The allowed $f_0$'s can be fully characterized by $H$-SPTs according to the results of Ref.~\cite{else2021goldstone}.

\subsection{The non-invertible case}\label{subsec:noninvertible}

We discuss the mathematical classification of non-invertible families in Sec.~\ref{sec:MainRes-noninv}. But it turns out some results in the non-invertible case can be obtained as a simple corollary of the results in the invertible case, as we now describe.
For the same reasons as discussed in the previous subsection, to get meaningful results we need to consider an appropriate quotient of the $H_0$-families. Specifically, the objects we study will be elements of the quotient $\{ \text{$H_0$-families over $\Lambda$} \} / \sim$, where we mod out by the equivalence relation $\sim$, where $f \sim f'$ if $f$ and $f'$ differ at most by stacking a constant invertible $H_0$-family over $\Lambda$. We want to find the compatibility relation between these objects and $G$-anomalies. For the rest of this subsection, for brevity when we refer to ``families'' we will actually mean elements of the aforementioned quotient.

Let us consider an anomaly $\alpha \in \{ \text{$G$-inv}\}_{d+1}^H$, i.e.\ a $G$-anomaly that becomes trivial upon restricting to $H$.
We make use of the result described in the previous subsection  that for any $\alpha \in \{\text{$G$-inv} \}^H_{d+1}$, there exists at least one $p_\alpha  \in \{ \text{$H_0$-pumps over $\Lambda$} \}$ such that $p_\alpha \tl \alpha$. 
Hence, given any system (not necessarily with invertible ground states) in which $G$ has anomaly $\alpha$ and is spontaneously broken to $H$, there is another system with \emph{invertible} ground states in which $G$ has anomaly $-\alpha$ and is spontaneously broken to $H$. If we take the stack of these two systems, then we obtain a system with trivial $G$-anomaly. Recall that the families that are compatible with the \emph{trivial} $G$-anomaly can be derived from the framework of Ref.~\cite{else2021goldstone} and correspond to those that can be obtained from the $H$-SPT (or, now that we are allowing non-invertible ground states, $H$-S\emph{E}T) carried by the symmetry-breaking ground states. To obtain the $H_0$-families over $\Lambda$ compatible with $G$-anomaly $\alpha$, one simply stacks these with the pump $p_\alpha$. In other words, the $G$-anomaly at most contributes an $H_0$-pump compared to the non-anomalous case.

However, this is not the complete story for the non-invertible case. Unlike for invertible families, it is not necessary for the anomaly to be trivial when restricted to $H$, because topologically ordered ground states can be $H$-symmetric even in the presence of a non-trivial $H$ anomaly. We will make some comments about how to proceed in this case in Sec.~\ref{sec:MainRes-noninv}.

\section{Examples}\label{sec:Apps}
\label{sec;Apps}

In this section we will illustrate the formalism discussed above through several examples, which we will analyze using physical arguments. Later, in Sections \ref{sec:GC-Classif} and \ref{sec:MainTechRes} we will explain how to compute the classification of families and give more systematic and abstract ways to understand the compatibility relation.

\subsection{Thouless pump}\label{sec:GAnomTrivialFam}

Consider fermionic systems with symmetry $G = \U(1)_a \times \U(1)_b$. Let us spontaneously break $\U(1)_a$ down to the trivial group but preserve $
\U(1)_b$. For example, $\U(1)_b$ can correspond to charge conservation, and contains the fermion parity subgroup, while $\U(1)_a$ is an emergent symmetry that appears in a low-energy description. A concrete instance of this is when we compactify a 2d integer quantum Hall state onto a cylinder with open boundaries. We get two gapless edge states with opposite chiralities at the ends, whose low-energy description is a Dirac fermion in (1+1)D. This theory has a $\U(1)_L \times \U(1)_R$ symmetry associated to individual charge conservation for the two chiral edges, with conserved quantities $N_L, N_R$. The $\U(1)_b$ and $\U(1)_a$ symmetries defined above correspond to the conservation of $N_L + N_R$ and the charge on any one edge, say $N_L$.

Let us fix $d=1$, where we have a $\Z$ classification of families over $\Lambda = S^1$ corresponding to the Thouless charge pumps. The pump is described by an effective theory $\frac{m}{2\pi} d\phi \wedge b$, where $\phi$ is the $S^1$ order parameter and $m$ is an integer. This effective action has two consequences. First, winding the order parameter as a function of time (so that $\int dt \partial_t \phi = 2\pi$) gives rise to a current of $b$ that pumps $m$ units of charge across the 1d system. Secondly, inserting a defect by changing the order parameter configuration in space (so that the winding number $\int dx \partial_x \phi = 2\pi$)
 causes the system to carry $m$ extra units of charge compared with the configuration where the order parameter is constant and the winding number is zero.

We now argue that the pumping invariant is compatible with, and requires, a mixed anomaly of $\U(1)_a$ and $\U(1)_b$. The anomaly can be defined as the failure of conservation of $\U(1)_a$ charge in the presence of an electric field of $\U(1)_b$:
\begin{equation}\label{eq:Thouless-anomaly}
    \partial_t n_a + \partial_x j_a = \frac{m}{2\pi} E_b.
\end{equation}
Here $m$ is an integer-quantized anomaly coefficient. Equivalently, the anomaly is the statement that the local charge densities $\hat{n}_a$ and $\hat{n}_b$ fail to commute: their commutation relation is
\begin{equation}\label{eq:Thouless-commutation}
    [\hat{n}_b(x),\hat{n}_a(y)] = -\frac{i m}{2\pi} \delta'(x-y).
\end{equation}
See Appendix C of Ref.~\cite{Else2021drag} for a derivation of Eq.~\eqref{eq:Thouless-commutation} from Eq.~\eqref{eq:Thouless-anomaly}. Let us now conjugate the anomalous 1d system with the many-body polarization operator associated to $\U(1)_a$:
\begin{equation}
    e^{i\hat{\Pi}} := e^{\frac{2\pi i}{L} \int dx' x' \hat{n}_a(x')}.
\end{equation}
$e^{i \hat{\Pi}}$ inserts a vortex of the $S^1$ order parameter  associated to $\U(1)_a$ symmetry breaking, since it is a large gauge transformation. At the same time, the total charge $\hat{N}_b := \int dx' n_b (x')$ does not commute with the polarization operator. In fact Eq.~\eqref{eq:Thouless-commutation} implies that
\begin{equation}
    e^{i \hat{\Pi}} N_b e^{- i \hat{\Pi}} = N_b + m.
\end{equation}
The polarization operator simultaneously inserts a vortex of the order parameter and (by virtue of the anomaly) changes the total $\U(1)_b$ charge by $m$. Hence, we conclude that the vortex carries charge $m$. This is a defining property of the Thouless pump of order $m$, and so we have shown that a mixed anomaly is equivalent to obtaining the Thouless pump upon spontaneously breaking $\U(1)_a$.

Note that the $G$ symmetry actually admits a $\Z^3$ classification of 1d anomalies. In addition to the mixed anomaly there are also anomalies of the individual $\U(1)$ symmetries. A (2+1) dimensional action for the fermionic invertible phase which cures the anomaly is 
\begin{equation}
     \mathcal{L}_{\text{anom}} = \frac{n_1}{4\pi} a \wedge da +  \frac{n_2}{2\pi} a \wedge db + \frac{n_3}{4\pi} b \wedge db,
\end{equation}
where we denote background gauge fields for the two $\mathrm{U}(1)$ symmetries as $a$ and $b$ respectively

The coefficient $m$ in the above discussion corresponds to $n_2$ in this action. The first part of the compatibility relation just states that the anomaly term with coefficient $n_2$ is compatible with a Thouless pump with pumping coefficient $n_2$. Let us now consider the compatibility relation for the full anomaly.

The coefficient $n_3 \in \Z$ classifies $\U(1)_b$ anomalies; these remain nontrivial with respect to the unbroken subgroup $H = \U(1)_b$, so we should set $n_3 = 0$ to get a gapped ground state. Also, $n_1 \neq 0$ is compatible with the trivial pump. Since the anomaly term corresponding to $n_2$ is $\U(1)_b$ independent, it can be realized in 
a system in which $\U(1)_b$ acts completely trivially, in which case we certainly can only have the trivial $\U(1)_b$ pump.

These facts, together with the linearity of the compatibility condition, imply that the full compatibility condition that relates the $\mathbb{Z}$-valued $\U(1)_b$ pump index to the $\mathbb{Z}^3$-valued $\U(1)_a \times \U(1)_b$ anomaly is
\begin{equation}\label{eq:Thouless-CR}
    m \tl (n_1, n_2, n_3) \quad \Leftrightarrow \quad m = n_2 \text{ and } n_3 = 0.
\end{equation} This example belongs to the case discussed in Sec.~\ref{subsec:simple_anomaly}. As predicted there on general grounds, we can define a homomorphism $\varphi$ from the anomalies such that the unbroken subgroup $H = \U(1)_b$ is anomaly free, into the pumps. Specifically, this homomorphism sends $ (n_1,n_2,0) \mapsto n_2$.

\subsection{Fermion parity pump on the boundary of quantum spin Hall state}\label{sec:QSH}

Let us consider a system in $d=1$ with symmetry $G = (\U(1)^f \rtimes \Z_4^{Tf})/\Z_2^f$ (which lies in Class AII of the free fermion symmetry classification). The bosonic symmetry is $G_b = \U(1)_b \rtimes \Z_2^T$, where $\U(1)_b := \U(1)^f/\Z_2^f$. The fermion carries charge 1/2 under $\U(1)_b$ and has a Kramers degeneracy.

Suppose we spontaneously break the $\U(1)^f$ symmetry down to $\Z_2^f$ fermion parity, where the residual symmetry is $H = \Z_4^{Tf}$ (which would place the boundary state in Class DIII). In this case, we can check that $H_0 = \Z_2^f$. The $d = 1$ $H_0$-pumps over $S^1$ are $\Z_2$ classified: inserting a vortex of the $S^1$ order parameter in space changes the fermion parity of the ground state. We refer to this as a fermion parity pump. Meanwhile, the $G$ anomalies are also $\mathbb{Z}_2$ classified; the non-trivial anomaly corresponds to being at the boundary of a ``Quantum Spin Hall'' (QSH) topological insulator in $d=2$.

We can understand the compatibility relation using the following argument. Let $[\alpha]$ and $[p]$ denote the QSH anomaly and the non-trivial pump. Let $[0]$ denote the trivial anomaly as well as the trivial pump. Then the full compatibility relation is
\begin{align}
    [p] &\tl [\alpha];  \quad [0] \cancel{\tl} [\alpha] \nonumber \\
    [0] &\tl [0];  \quad[p] \cancel{\tl} [0].
\end{align}
To intuitively see the first two relations, 
we use the defining property of the QSH insulator, namely that the fermion parity of the ground state is different depending on whether it has 0 or $\pi$ flux of $\U(1)^f$.

Suppose the system is on a spatial ring, and the winding number of the order parameter around the ring is $W$. Then by making a $\U(1)^f$ gauge transformation\footnote{We are allowed to perform such gauge transformations however we like because $\U(1)^f$ is not itself anomalous in the case of the QSH anomaly.}, we can relate this to a system where the order parameter has zero winding number, but the system caries flux $W \pi$ of $\U(1)^f$.
The boundary anomaly of the QSH state implies that the ground state in the presence of a $\pi$ flux carries fermion parity $-1$ compared to the zero flux ground state. Therefore, we conclude that an elementary vortex (winding number of the order parameter around the ring is 1) carries fermion parity $-1$ relative to the vortex-free ground state, which is the signature of the non-trivial fermion pump $[p]$.
From this we conclude that $[p] \tl [\alpha]$, and also that $[0] \cancel{\tl} [\alpha]$.

Now the relation $[0] \tl [0]$ always holds, and these three properties are sufficient to also enforce that $[p] \cancel{\tl} [0]$. However, let us give a complementary way of understanding this fact. If we just have $G = \U(1)^f$ and $H = \Z_2^f$, we can in fact realize a fermion pump $[p]$ with $[p] \tl [0]$. Specifically, for a 1D system without any bulk, we construct a family such that the ground state can be adiabatically connected to a Kitaev chain for any value of the order parameter. By the argument above, the properties of a vortex are related to the properties of a $\pi$ flux -- but it is well-known that threading a fermion parity flux through a ring carrying a Kitaev chain changes the fermion parity of the ground state. However, the Kitaev chain is not symmetric under $\Z_4^{Tf}$, so the possibility of $[p] \tl [0]$ is forbidden once we impose $\Z_4^{Tf}$. (These considerations also show why the QSH anomaly is actually equivalent to the trivial anomaly once the time-reversal symmetry is broken.)

Note that there are in fact two distinct families $[f_1], [f_2]$ which correspond to fermion parity pumps but differ in their constant part; both are compatible with $[\alpha]$. Their difference $[\delta] = [f_1] \otimes [f_2]^{-1}$ under stacking should be equivalent to a non-trivial Class DIII topological superconductor, which generates a $\Z_2$ classification of $H$-SPT states in $d=1$. This is a consequence of the general theory developed in Sec.~\ref{sec:Summary}. 

In a derivation shown in Sec.~\ref{app:QSH} we apply a mathematical framework that reproduces the above results.

\subsection{Superconducting proximity effect on the surface of a 3d topological insulator}

We can similarly ask about breaking $\U(1)$ symmetry on the surface of a 3d topological insulator, for which $G, H, H_0, \Lambda$ are the same as in the previous example. This can be achieved by placing the surface in proximity to an s-wave superconductor. In this case, we have the famous result of Fu and Kane \cite{Fu2008} that the proximity effect gives rise to surface states which preserve time reversal symmetry, but resemble a $p+ip$ superconductor in that a vortex of the superconducting order parameter carries an unpaired Majorana zero mode. This corresponds to the non-trivial element of the $\mathbb{Z}_2$ classification of pumps over $S^1$ with only fermion parity symmetry, i.e.\ the ``pump of Kitaev chain''.
Thus, the non-trivial anomaly is compatible with the non-trivial pump, but not the trivial pump.
Meanwhile, one can show that the trivial anomaly is  compatible only with the trivial pump, as the non-trivial pump would require the 2D system to have a $p+ip$ superconductor ground state, which is disallowed by time-reversal symmetry.

\subsection{Non-invertible family with anyon permutation}\label{sec:AnyonPerm}

Let us now consider a situation in which a $\U(1)$ subgroup of $G$ is broken so that the resulting family over $S^1$ is topologically ordered, and specifically has the property that cycling the $S^1$ leads to a permutation of the ground states. To be concrete, we can consider the topological order $\mathcal{C}$ to be that of the $\Z_2$ gauge theory describing the toric code, with Abelian anyons $1, e, m, \psi = e \times m$ forming a $\Z_2 \times \Z_2$ group under fusion. The anyons $e,m$ are bosons, while $\psi$ is a fermion. Now there is a $\Z_2$ automorphism of $\mathcal{C}$ which interchanges $e$ and $m$. The nontrivial family mentioned above then has the property that a vortex of the order parameter localizes a $\Z_2$ twist defect, so that an $e$ particle going around the vortex gets permuted to $m$, and vice versa.

One way to obtain this family is to require that the original $G = \U(1)$-symmetric state has trivial anomaly, and $G$ spontaneously breaks down to $H = \Z_2$, so that a vortex of the $S^1$ order parameter corresponds to a flux of the residual $\Z_2$ symmetry. Therefore, if the ground states are in an SET phase with respect to this residual symmetry, characterized by the fact that the $\Z_2$ symmetry is anyon-permuting, then a vortex of the order parameter will permute the anyons in the same way as a $\Z_2$ flux.

Now we can ask if such an anyon-permuting family can arise if the $\U(1)$ symmetry has anomaly and is completely broken (i.e.\ $H$ is trivial). The answer is no, because from Sec.~\ref{subsec:noninvertible} every topologically ordered family can be written in terms of an $H$-SET invariant and the invariant of some invertible family. When $H$ is trivial, the associated SET family invariant must be trivial. Moreover, invertible families by definition cannot permute anyons. Therefore, anyon-permuting families cannot arise due to a $G$-anomaly; they are always associated to an $H$-SET invariant, where $H$ activates the anyon permutation.

\subsection{Quantum Hall ferromagnet}\label{sec:QHFM}
A quantum Hall ferromagnet is a 2d electron system with a $G = \U(2)$ symmetry, which corresponds to rotations in either spin or valley space; the latter applies for example in graphene-based systems. This symmetry can be spontaneously broken to a subgroup $H = \U(1)_{\uparrow} \times \U(1)_{\downarrow}$ by strong interactions such as antiferromagnetic order. Here `up', $\uparrow$ and `down', $\downarrow$ refer to isospins aligned along some direction, and the diagonal $\U(1)$ subgroup corresponds to total number conservation, which we will denote $\U(1)_c$. We will assume that the symmetry-broken ground state is fully spin-polarized, so that the down spins states, say, are empty, while only the up spin states are filled. 

In terms of our usual notation, the set of possible isospin orientations spans a space $\Lambda = G/H = S^2$. We have $H_0 = \U(1)_c$ (total number conservation).

\subsubsection{Integer case}

First we consider the case of an invertible family over $S^2$. This was discussed previously in Ref.~\cite{else2021goldstone}, but we will repeat the main details here to illustrate the formalism. The basic property of the family is that an $S^2$ order parameter defect, which is a skyrmion, carries unit charge under $H_0$. This is encoded by the following response action:
\begin{equation}\label{eq:QHFM}
    S = S_0 + \frac{1}{4\pi} \int d^3 x \epsilon^{\mu \nu \lambda} \epsilon^{a b c} A_{\mu} n_a \partial_{\nu} n_b \partial_{\lambda} n_c .
\end{equation}
Here $A$ is a background gauge field for the $\U(1)_c$ symmetry, while $\vec{n} \in S^2$ is an order parameter field. We will now show that this effective action can be derived from that of a spin-polarized IQH state, establishing that the pump is determined by an $H$-SPT.

An $S^2$ skyrmion can be viewed as a sum of a $2\pi$ flux of $\U(1)_{\uparrow}$ and a $-2\pi$ flux of $\U(1)_{\downarrow}$ \cite{else2021goldstone}. This can be seen from the homotopy exact sequence for the fiber bundle $H \rightarrow G \rightarrow \Lambda$ with $G = \U(2), H = \U(1) \times \U(1)$:
\begin{equation}
1 \rightarrow   \pi_2(S^2) = \Z \xrightarrow{n \rightarrow (n,-n)} \pi_1(H) = \Z_{\uparrow} \times \Z_{\downarrow} \xrightarrow{(a,b) \rightarrow a + b} \pi_1(\U(2)) = \Z \rightarrow 1
\end{equation}
This implies that we have a map
\begin{equation}
    H^2(BH,\Z) = \Z \times \Z \xrightarrow{(a,b) \rightarrow a-b} \rightarrow H^2(S^2,\Z) = \Z.
\end{equation}
The left-hand side are $d=0$ SPTs of $H$, whose effective actions are generated by $dA_{\uparrow}, dA_{\downarrow}$. The generator of the right-hand side is the skyrmion density $\epsilon^{a b c} n_a \partial_{\mu} n_b \partial_{\nu} n_c$. The map states we can replace $(dA_{\uparrow})_{\mu \nu}$ and $-(dA_{\downarrow})_{\mu \nu}$ with the skyrmion density.  $dA_{\uparrow} - d A_{\downarrow}$. We can also restrict any effective action for $H$ to $H_0$ by replacing $A_{\uparrow} = A_{\downarrow} = A$, where $A$ is a gauge field for $H_0$. Let us now start with an IQH state for the $\uparrow$ spins, with action $S_{\uparrow} = S_0 + \frac{1}{4\pi}\int d^3 x A_{\uparrow} \wedge dA_{\uparrow}$. The above replacements then directly give Eq.~\eqref{eq:QHFM}.

We can recover the same understanding from the general theory. There are no $\U(2)$ anomalies in $d=2$, while the pump (which is the quantum Hall ferromagnet) is $\Z$ classified. The full classification of families is $\Z \times \Z$, where the second factor corresponds to IQH states. The compatibility relation simply states that each pump is compatible with the trivial anomaly.

Finally, $d=2$ SPT states with $H$ symmetry are $\Z^3$ classified, corresponding to the three CS terms $\frac{k_1}{4\pi} A_{\uparrow} d A_{\uparrow}, \frac{k_2}{2\pi} A_{\uparrow} d A_{\downarrow}, \frac{k_3}{2\pi} A_{\downarrow} d A_{\downarrow}$. From the general derivation of Appendix ~\ref{app:QHFM} we get a surjective homomorphism $\Z^3 \rightarrow \Z^2$ which determines each family in terms of an $H$-SPT, as we explicitly showed above.

\subsubsection{Fractional case}

We can also consider \textit{fractional} quantum Hall ferromagnets, in which strong interactions drive the isospin degrees of freedom into a topologically ordered state. In this case, the skyrmions can acquire fractional charges under $H_0$; this is argued as follows. Suppose the FQH state is completely spin-polarized with spin $\uparrow$. The basic property of symmetry fractionalization is that a $2\pi$ flux of $\U(1)_{\uparrow(\downarrow)}$ induces some Abelian anyon $a_{\uparrow (\downarrow)}$ in the topological order. Therefore, an elementary skyrmion induces the anyon $a = a_{\uparrow}$ (there is no symmetry fractionalization in the $\downarrow$ sector since in a fully spin-polarized state $\U(1)_{\downarrow}$ acts trivially). But $a$ carries a fractional charge $Q_a$ under $H_0 = \U(1)_c$, as given by the braiding statistics relation $e^{2\pi i Q_a} = M_{v,a}$. Here $v$ is the Abelian anyon induced by inserting $2\pi$ flux of $\U(1)_c$, and corresponds to the `charge vector' in a $K$-matrix CS description of the FQH state. As a result, skyrmions also carry charge $Q_{a} \mod 1$. Note that in this case, the non-invertible family over $S^2$ is described by an SET invariant associated to the $H$ symmetry, namely the symmetry fractionalization data that fixes the anyons $a$ and $v$. 

Let us consider some examples. Suppose the topological order is described by a two-component CS theory with the following $K$-matrix and charge vector:
\begin{equation}
    K = \begin{pmatrix}
        m_1 & n \\ n & m_2
    \end{pmatrix}, \quad q = \begin{pmatrix}
        1 \\ 1
    \end{pmatrix}.
\end{equation}
This $K$ matrix describes the general Halperin $(m_1,m_2,n)$ state that can be realized in the presence of suitable interlayer interactions. These states occur at filling $\nu = \frac{m_1 + m_2 - 2n}{m_1 m_2 - n^2}$. For a fully spin-polarized state with only $\uparrow$ spins filled, we require the filling $\nu_{\downarrow} = \frac{m_1 - n}{m_1m_2 - n^2}$ to vanish. Therefore we take $m_1 = n$. Now since a skyrmion is identified with a $2\pi$ flux of $A_{\uparrow}$ and a $-2\pi$ flux of $A_{\downarrow}$, it induces the anyon $a = (1,-1)^T$. Therefore the charge of the skyrmion is given by $Q_{a} = q^T K^{-1} a = \frac{m_2 - m_1}{m_1 m_2 - n^2} \mod 1$.

From a more general perspective, the symmetry fractionalization in the $H$-SET is given by an element of $H^2(BH,\mathcal{A})$ where $\mathcal{A}$ is the group of Abelian anyons. But using the map $\pi_2(S^2) \rightarrow \pi_1(H)$, we obtain a map $\HH^2(H,\mathcal{A}) \rightarrow H^2(S^2,\mathcal{A})$, which takes $(a_{\uparrow},a_{\downarrow}) \rightarrow a_{\uparrow}$ in our case.\footnote{The Hurewicz theorem shows that $\pi_2(S^2) \cong H_2(S^2,\Z)$ and $\pi_1(\U(1)) \cong \HH_2(\U(1),\Z)$ ($H_2$ and $\HH_2$ denote the singular and group homology). Using the injective map $\pi_2(S^2) \rightarrow \pi_1(H)$, we get a dual map $\text{Hom}(H_2(H,\Z),\Z) = \HH^2(H,\Z) = \Z \times \Z \rightarrow \text{Hom}(H_2(S^2,\Z),\Z) = H^2(S^2,\Z) = \Z$.  Finally, we use the facts that $\HH^2(H,\mathcal{A}) = \HH^2(H,\Z) \otimes \mathcal{A}$ and $H^2(S^2,\mathcal{A}) = H^2(S^2,\Z) \otimes \mathcal{A}$, which follows from the Universal Coefficient Theorem for cohomology.} Therefore, the symmetry fractionalization in the $H$-SET directly determines the fractional charge induced by an $S^2$ skyrmion.

\subsection{$\SU(2)$ symmetry breaking in $d=1$ and higher Berry curvature}\label{sec:HBC}

Consider systems with $G = \SU(2)$ in $d=1$, whose anomalies are classified by $\HH^3(\SU(2),\U(1)) = \Z$. The anomaly can be described by a Chern-Simons term at level $k \in \Z$ for an $\SU(2)$ gauge field $a_{\mu}$ on a 3-dimensional space-time manifold $M$:
\begin{equation}
    S_{CS} = \frac{k}{4\pi} \int_M d^3 x \epsilon^{\mu \nu \lambda} \text{Tr} \left( a_{\mu} \partial_{\nu} a_{\lambda} + \frac{2}{3} a_{\mu} a_{\nu} a_{\lambda} \right).
\end{equation}
Note that $\SU(2)$ as a topological space is homotopy equivalent to $S^3$, and therefore it is possible to break $G$ completely down to $H = \Z_1$ via a family of mass terms parameterized by $S^3$. Although spontaneous symmetry breaking in $d=1$ is not allowed when we incorporate dynamical fluctuations of $a_{\mu}$, for the present analysis we can essentially treat $a_{\mu}$ as a static background field, so it is still meaningful to talk about Goldstone modes and their effective actions. 

It has long been known (see for example \cite{Elitzur1989CS}) that breaking $\SU(2)$ on the boundary of $M$ leads to a non-linear sigma model containing the following Wess-Zumino-Witten term, where $\partial M$ contains the physical $(1+1)D$ system of interest and the action does not depend on the choice of extension to $M$:
\begin{equation}
    S_{\text{WZW}} = \frac{k}{12\pi}  \int\limits_{M} \epsilon^{\mu \nu \lambda} \text{Tr } [g^{-1} \partial_{\mu} g g^{-1} \partial_{\nu} g  g^{-1} \partial_{\lambda} g ]
\end{equation}

The above response corresponds to the `higher Berry curvature' and associated higher Chern number studied in recent works \cite{Kapustin2020HBC,Wen2021HBC}. Since we are studying bosonic systems with $H$ trivial, the $H$-SPT classification in $d=1$ vanishes. The $G$-anomalies are classified by $H^4(B\SO(3),\Z) = \Z$, while the classification of families is given by $H^3(S^3,\Z) = \Z$. The compatibility relation in this case becomes an isomorphism $\varphi: \Z \rightarrow \Z$, which is consistent with the discussion in Sec.~\ref{subsec:simple_anomaly}. The mathematical computations for this example are given in Appendix~\ref{app:S3}.

\subsection{Chern number pump}\label{sec:Chern_pump}
Starting with the Berry-Chern number family in $d=0$, which is a family over $S^2$, one can construct a Chern number pump, which is a $d=1$ family over $\Lambda = S^2 \times S^1$ \cite{Wen2021HBC} in which an $S^1$ order parameter defect localizes a $d=0$ family characterized by a nontrivial Berry-Chern number. (As discussed in Ref.~\cite{Wen2021HBC}, a slight modification to this construction gives the family over $S^3$ mentioned above.) We can ask if the Chern number pump can be realized by spontaneous symmetry breaking, from a state with $G$ symmetry. Suppose we take $G = \U(1) \times \SO(3)$, and assume that the $\U(1)$ subgroup is completely broken to give an $S^1$, while the $\SO(3)$ subgroup is broken down to $\SO(2)$. Here we have $H = \SO(2)$ and $H_0 = \Z_1$. 

In this case, a calculation shown in Sec.~\ref{app:Chern_pump} tells us that the Chern number pumps are not compatible with any $G$ anomaly. Indeed, there are no mixed anomalies of $\U(1) \times \SO(3)$ in $d=1$, while the pure $\U(1)$ or $\SO(3)$ anomalies are not compatible with the Chern number pump.  Therefore, this Chern number pump cannot be realized by spontaneously breaking a $\U(1) \times \SO(3)$ symmetry, with or without anomaly.

Instead, it can be realized by starting from a $d=2$ family over $S^2$ with $\U(1)$ symmetry, rather than a $G$-SPT. The property of this family is that a skyrmion of the $S^2$ traps a quantized $\U(1)$ charge. We can break $\U(1)$ symmetry on the boundary of the family to get the Chern number pump, but this is different from the usual setup of spontaneous symmetry breaking followed in this paper.

The phenomenon of a pump being compatible with a family in one higher dimension rather than a $G$ anomaly is a common feature in systems where we start with a symmetry $G = G_1 \times G_2$ and break symmetry via mass terms that span a product space $\Lambda = \Lambda_1 \times \Lambda_2$. To take another example, consider bosonic systems with $G = \U(1) \times \U(1) \times H$ with $\Lambda_1 = \Lambda_2 = S^1$. The pumps in dimension $d$ which become trivial if we set either $\Lambda_1$ or $\Lambda_2$ to a constant value are captured by $H^1(S^1,H^1(S^1,H^{d}(BH,\Z))) \cong H^{d}(BH,\Z)$. On the other hand, the mixed anomalies of $\U(1) \times \U(1) \times H$ in dimension $d$ are classified by a subgroup of $H^{d+3}(G,\Z)$ which is $H^2(B\U(1),H^2(B\U(1),H^{d-1}(BH,\Z))) \cong H^{d-1}(BH,\Z)$. This dimension mismatch makes it evident that the pump cannot be compatible with a $G$-anomaly.

\section{Generalized cohomology classification of invertible states and families}\label{sec:GC-Classif}
 In this section we will describe a mathematical formalism that allows us to give a general characterization of the compatibility relation that we set up in Section \ref{sec:Summary}. It will also form the necessary background for the spectral sequence approach that we develop in Sec.~\ref{sec:MainTechRes} below.

\subsection{Notation and general classification picture}\label{sec:MainRes-TrivAnom}

Below we review some material that has appeared in various prior works, see for example Refs.~\cite{Kitaev2006,Thorngren2018}. 

Let $\Theta_d$ be the space of gapped ground states in $d$ space dimensions without symmetry. In the invertible case we can just write $\Theta_d$, while in the case where these ground states have topological order corresponding to some anyon theory $\mathcal{C}$, we can write this as $\Theta_d^{\mathcal{C}}$. If we only consider ground states invariant under a global $H_0$ symmetry, we denote the corresponding spaces of states as $\Theta_d^{H_0}$ and $\Theta_d^{\mathcal{C};H_0}$ respectively. While the statements below are equally applicable to invertible and topologically ordered states, for convenience we will drop the $\mathcal{C}$ superscript for the rest of this section. We will assume that all the symmetries act unitarily, and in the case of fermionic systems, we will assume that the fermionic symmetry group is of the form $G_f = G \times \Z_2^f$, where $\Z_2^f$ is fermion parity\footnote{\label{footnote:more_general}We can also consider the more general case where the symmetry group is a direct product $G \times K$, where all the anti-unitary symmetries, as well as the fermion parity (in the fermionic case) are elements of $K$, provided that we do not try to break $K$. Then we can simply replace $\Theta_d$ with $\Theta_d^{K}$ the space of bosonic/fermionic states invariant under $K$, and we define $H$, $H_0$ as subgroups of $G$ rather than $G \times K$.}.

The general classification picture can be summarized by the following three points:
\begin{enumerate}
    \item Topological phases in $d$ dimensions with symmetry $H_0$ correspond to homotopy classes of maps from $BH_0 \rightarrow \Theta_d$, where $BH_0$ denotes the classifying space of $H_0$.
    \item \textit{Families} of gapped ground states in $d$ dimensions with symmetry $H_0$ over a space $\Lambda$ (on which $H_0$ acts trivially) correspond to homotopy classes of maps from $\Lambda \times BH_0 \rightarrow \Theta_d$. This includes the case where the family is obtained by spontaneously breaking an anomalous $G$ symmetry.
    \item In the case where there is a \textit{non-anomalous} symmetry $G$ which is spontaneously broken to $H$, and $\Lambda = G/H$, the resulting families of gapped ground states with symmetry $H_0$ correspond to homotopy classes of maps from $\Lambda//G \rightarrow \Theta_d$, where $\Lambda//G$ is called the homotopy quotient of $\Lambda$ by its $G$ action, and is defined below.
\end{enumerate}

First, note that for any group $G$, $BG \cong EG/G$, where $EG$ is a weakly contractible space (i.e. having trivial homotopy groups) on which $G$ acts freely (without any fixed points). Maps from a manifold $M$ to $BH_0$ correspond to $H_0$ gauge fields on $M$, and so maps from $BH_0 \rightarrow \Theta_d$ essentially encode all the distinct ways in which the low-energy effective theory can be coupled to a background $H_0$ gauge field. In the invertible case, these maps correspond to the decorated domain wall construction of invertible states, as described in Sec.~\ref{sec:Inv-classif}, while in the topologically ordered case, they encode the symmetry fractionalization and topological response of the system, as described for example in Ref.~\cite{barkeshli2019}. 
  
Now suppose we have an order parameter valued in $\Lambda$, in addition to the $H_0$ symmetry. If $H_0$ is trivial, the classification of families is clearly given by homotopy classes of maps from $\Lambda \rightarrow \Theta_d$ (since for each value $\lambda \in \Lambda$ of the order parameter we have a gapped ground state $\ket{\Psi_{\lambda}} \in \Theta_d$). This agrees with the intuition that the topological response of the family on a manifold $M$ should be expressed in terms of effective actions that are functionals of order parameter fields, i.e.\ of maps $M \rightarrow \Lambda$. If we include an $H_0$ symmetry which does not act on $\Lambda$, we need to consider the order parameter field together with an $H_0$ gauge field, therefore we replace the space $\Lambda$ with $\Lambda \times BH_0$.

Finally, in the case where there is a non-anomalous symmetry $G$ which acts non-trivially on $\Lambda$, the response of the family is determined by its coupling to a background $G$ gauge field together with an order parameter field that has a $G$ action. This combined object needs to be equivariant under the $G$ action: for example, an order parameter field configuration $\lambda(x)$ can be shifted to $g \lambda(x)$ for $g \in G$ without changing the definition of the background $G$ gauge field, but $G$ gauge transformations will generically change $\lambda(x)$. It turns out to be appropriate to define this object in terms of maps from $M$ to a space $\Lambda//G := (EG \times_G \Lambda)/G$. The space $EG \times_G \Lambda$ is homotopy equivalent to $\Lambda$ (since $EG$ has vanishing homotopy groups), but is also acted upon freely by $G$ (since $G$ acts freely on $EG$). The symbol $\times_G$ denotes the diagonal action $g(\lambda, e) := (g \lambda, g^{-1} e), \lambda \in \Lambda, e \in EG$. Note that $\Lambda//G$ reduces to $BG$ when $\Lambda$ is trivial, to $\Lambda \times BG$ when $G$ acts trivially on $\Lambda$, and to $\Lambda$ when $G$ is trivial.

Note that when $\Lambda = G/H$, $\Lambda//G$ is actually homotopy equivalent to $BH$, the classifying space of $H$. A proof of this is given in Ref.~\cite{else2021goldstone}. Thus the families compatible with trivial $G$-anomaly are classified by maps from $BH \rightarrow \Theta_d$; they are in one-to-one correspondence with $H$-symmetric topological states. In fact, we can relate the second and third classifications above using the projection map from $\Lambda \times BH_0 \rightarrow \Lambda//G$, which amounts to identifying points in $\Lambda \times BH_0$ that are equivalent under the $G$ action. Given a map from $\Lambda//G \rightarrow \Theta_d$, we can therefore obtain a map from $\Lambda \times BH_0 \rightarrow \Theta_d$. In the invertible case, this tells us that families associated to breaking a non-anomalous symmetry form a normal subgroup of the classification of all families.

All the above arguments are unaffected by replacing $\Theta_d$ with $\Theta_d^{\mathcal{C}}$, for some $\mathcal{C}$. Therefore it applies equally to topologically ordered families, if instead of SPT/invertible states we consider SET states.

\subsection{Review of generalized cohomology hypothesis} \label{sec:Inv-classif}
Each classification discussed above is defined via homotopy classes of maps from a suitable space into $\Theta_d$. Let us now specialize to the case of invertible states. If we make the assumption that the spaces $\Theta_d$ form an $\Omega$-spectrum, we can explicitly compute these classifications in terms of a generalized cohomology theory. Below we will briefly review the main arguments involved, which have been discussed in several prior works; see for example Refs.~\cite{Gaiotto:2017SPT,Xiong:2016deb,wang2021domain}.

A set of spaces $\Theta_d, d \in \Z$, is said to form an $\Omega$-spectrum if it satisfies the following conditions: each $\Theta_d$ has a basepoint 0 (which can be interpreted as the trivial product state in dimension $d$), and if $\Omega X$ denotes the space of based loops in $X$, then $\Omega \Theta_{d+1}$ is homotopy equivalent to $\Theta_d$. This implies that $\pi_k(\Theta_d) \cong \pi_{k+1}(\Theta_{d+1})$ for each $k$ and $d$. 

Given an $\Omega$-spectrum $\Theta = \{\Theta_d\}$. We define the \textit{generalized cohomology} $h^{d+1}(X)$ of a topological space $X$ as the group of homotopy classes of maps from $X\rightarrow \Theta_d$.  We claim that elements of this group classify invertible states or families with symmetry in space dimension $d$, depending on the choice of $X$. 

First assume $X$ is a point. The group $h^{d+1}(\text{pt})= \pi_0(\Theta_d)$ gives the classification of invertible phases without symmetry. In bosonic systems $h^{d+1}(\pt)$ is trivial in $d=0,1$, and $\Z$ in $d=2$ (this factor is generated by the invertible $E_8$ state). In the fermionic case, $h^{\bullet}(\text{pt})$ gives the classification of phases with with only $\Z_2^f$ fermion parity symmetry. Then $h^{d+1}(\text{pt})$ equals $\Z_2$ in $d=0$ (generated by the complex fermion), $\Z_2$ in $d=1$ (generated by the topological phase of the Kitaev chain) and $\Z$ in $d=2$ (generated by the chiral $p+ip$ superconductor). 

Next, suppose $X = BG$ for some $G$. In the bosonic case, if $G$ contains only unitary symmetries, the classification of phases is given by $h^{d+1}(BG)$. Similarly, in the fermionic case, if the fermionic symmetry group is a direct product $G_f = G_b \times \mathbb{Z}_2^f$ then the classification is given by $h^{d+1}(BG_b)$ (where $h$ is now the fermionic generalized cohomology theory).

\subsection{General homotopy-theory point of view on the compatibility relation}
It turns out that the compatibility relation that we introduced on physical grounds in Section \ref{sec:Summary} can be formulated entirely in the homotopy-theoretic point of view that we have described above. We give the details in Appendix \ref{app:Homotopy}. This formulation, however, is not in itself particularly amenable to concrete calculations. Instead, in the next section we will describe a perspective on the compatiblity relation in terms of spectral sequences.

\section{Compatibility spectral sequence for invertible families}\label{sec:MainTechRes}
The main goal of this section is to present a mathematical framework to explicitly compute the compatibility relation between a given $G$ anomaly and a given family over $\Lambda = G/H$ with symmetry $H_0$. First we review the decorated domain wall construction of invertible states and families using spectral sequences. Then we introduce a different object which we refer to as the `compatibility spectral sequence': this lets us interpolate between the classifications of $H_0$-families over $\Lambda$, $G$-anomalies, and $H$-invertible states, and gives an explicit way to compute the compatibility relation.

\subsection{Decorated domain wall construction}\label{subsec:DDW}

First we will briefly review the decorated domain wall construction of invertible states and invertible families; a more detailed explanation can be found, for example, in Refs.~\cite{Gaiotto:2017SPT,wang2021domain}. Since the compatibility relation is ultimately a relation between these different classifications, understanding the decorated domain wall construction is a useful preliminary step.  

Suppose there is a group $G$ with a normal subgroup $H_0$ whose classification is known, and we would like to compute $h^{d+1}(BG)$ using our knowledge of $h^{n}(BH_0)$ for different $n$. (While in most of this paper $H_0$ is a specific normal subgroup associated to our symmetry breaking problem, the mathematical results in this section still hold if we pick $H_0$ to be any normal subgroup of $G$.) Define $G^* := G/H_0$. Consider a network of $G^*$-defects in space-time dimension $d+1$. At each $k$-dimensional defect junction, we decorate $H_0$-invertible states. The full decorated domain wall state is defined by a set of decorations for each $k$. If we do not demand that decorations in different junctions be consistent with each other, the decorations at a $k$-dimensional junction are classified by $H^k(BG^*,h^{d+1-k}(BH_0))$. However, in general the decorations in dimension $k$ will constrain those in dimension $l<k$. A decorated domain wall state can be made $G$-symmetric and gapped if its constituent decorations in different dimensions satisfy all the constraints.

A systematic way to implement these constraints is through the Atiyah-Hirzebruch spectral sequence (AHSS), which is a general mathematical tool to compute the generalized cohomology of spectra.  In-depth discussions of the AHSS geared towards SPT classifications can be found in Ref.~\cite{wang2021domain}; for a standard mathematical treatment, see Ref.~\cite{Mccleary2000}. We will assume a working knowledge of spectral sequences in the rest of this section.

The decorations classified by $H^p(BG^*,h^q(BH_0))$ appear on the ``$E_2$-page" of the AHSS. A set of differentials encodes the constraints that these decorations need to satisfy in order to give a gapped ground state. Finally, the classification of $G$-symmetric states is obtained from the final set of consistent decorations by solving a sequence of group extentions. The AHSS therefore starts with the above $E_2$-page and returns the generalized cohomology of $BG$ after implementing the differentials and solving the extension problem. We say it converges to the cohomology of $BG$, and use the notation
\begin{equation}
\label{eq:lhs_generalization}
    E_2^{p,q}:= H^p(BG^*,h^q(BH_0)) \implies h^{p+q}(BG).
\end{equation}
Note that there are a number of variants of the AHSS which can be very useful in explicit computations, and are discussed at length in Ref.~\cite{Mccleary2000}. A simple variation is to consider different decompositions $H_0,G^*$ for a given $G$, but still relying on Eq.~\eqref{eq:lhs_generalization}.

A more general formulation of the AHSS that we will need is that if there is a fibration of topological spaces
$F \rightarrow E \rightarrow B$, and a spectrum $h^q(F)$ associated to the fiber $F$, there is an AHSS which converges to $h^*(E)$:
\begin{equation}\label{eq:AHSS-fibration}
    E_2^{p,q} := H^p(B,h^q(F)) \implies h^{p+q}(E).
\end{equation}
Here we do not require $F,E,B$ to correspond to the classifying space of some group. We can use this version with $B = \Lambda, F = BH_0, E = \Lambda \times BH_0$ and formulate an AHSS which converges to the classification of invertible families over $\Lambda$ with symmetry $H_0$:
\begin{equation}
    E_2^{p,q} := H^p(\Lambda,h^q(BH_0)) \implies h^{p+q}(\Lambda\times BH_0).
\end{equation}
This also has a decorated domain wall interpretation. On the $d$-dimensional domains, we decorate $d$-dimensional $H_0$-invertible states. Then on order parameter defects in spatial codimension $k>0$ we decorate $d-k$-dimensional $H_0$-invertible states. Again, a set of differentials specifies the way in which decorations in different codimensions can coexist.

To distinguish between the multiple spectral sequences that we will encounter in the next section, let us set up the following notation:
\begin{enumerate}
    \item We call the spectral sequence defined by $H^p(BG^*,h^q(BH_0)) \implies h^{p+q}(BG)$ the \textit{$G$-spectral sequence}. Its $p+q=d+2$ diagonal converges to the classification of $G$-anomalies in dimension $d$.
    \item Similarly, we call the spectral sequence defined by $H^p(BH^*,h^q(BH_0)) \implies h^{p+q}(BH)$ the \textit{$H$-spectral sequence}. Its $p+q=d+1$ diagonal converges to the classification of $H$-invertible states in dimension $d$.
    \item We call the spectral sequence defined by $H^p(B\Lambda,h^q(BH_0)) \implies h^{p+q}(\Lambda \times BH_0)$ the \textit{$\Lambda$-spectral sequence}. Its $p+q=d+1$ diagonal converges to the classification of invertible families over $\Lambda$ with symmetry $H_0$, in dimension $d$.
\end{enumerate}

\subsection{Derivation of compatibility relation}\label{sec:MainRes-inv}

In this section we finally derive the compatibility relation between an anomaly classified by $h^{d+2}(BG)$ and an invertible family classified by $h^{d+1}(\Lambda \times BH_0)$. In the case where the anomaly is trivial, we also show that the allowed family invariants are fully determined by $H$- invertible states classified by $h^{d+1}(BH)$. We saw above that spectral sequences appear naturally in the question of classifying invertible states and invertible families. This motivates us to address the compatibility question using a different but related spectral sequence that we term the `compatibility spectral sequence'. Note that the same computational method was used previously in Ref.~\cite{Lee2021WZW}, which analyzed Wess-Zumino-Witten terms for $\SU(n)$ and $\SO(n)$ symmetry breaking.  

As we described in Section \ref{sec:MainRes-TrivAnom}, we will focus on the case where the symmetries are unitary, and for fermionic symmetries we assume the fermionic symmetry takes the form $G_f = G \times \mathbb{Z}_2^f$. The main result we need is that for any normal subgroup $H_0$ of $H$, there is a fibration
\begin{equation}\label{eq:fibration}
    \Lambda \times BH_0 \rightarrow BH \rightarrow BG^* = B(G/H_0).
\end{equation}
A proof is sketched in Appendix~\ref{app:Proofs}. Using Eq.~\eqref{eq:AHSS-fibration}, we can form an AHSS for the above fibration converging to the cohomology of $BH$:
\begin{equation}
    E_2^{p,q} := H^p(BG^*,h^q(\Lambda\times BH_0)) \implies h^{p+q}(BH).
\end{equation}
Note that this AHSS is \textit{not} the same as the $G$ spectral sequence or the $\Lambda$ spectral sequence, although it has some terms in common with both of them. We refer to it as the \textit{compatibility spectral sequence}, or compatibility AHSS. The main difference between the compatibility AHSS and the other spectral sequences is that here we are decorating invertible \textit{families} on $G^*$ domain walls, while in the other cases we always decorate invertible states.

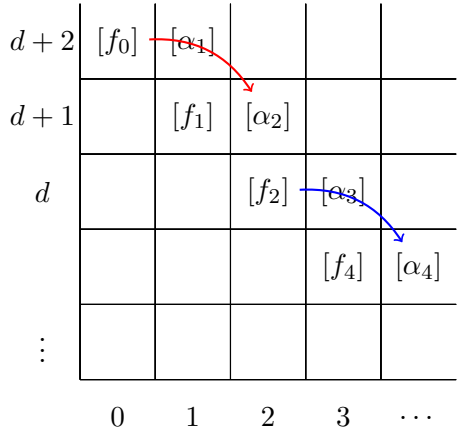
\begin{figure}
\centering
    
     \begin{tikzpicture}

\def\rows{5}
\def\cols{5}

\foreach \x in {0,1,2,3,4} {
    \foreach \y in {0,1,2,3,4} {
        \draw (\x,0) -- (\x,\rows);
        \draw (0,\y) -- (\cols,\y);
    }
}


\node at (1-0.5,-0.5) {0};
\node at (-0.5,1-0.5) {\vdots};
\node at (2-0.5,-0.5) {1};
\node at (3-0.5,-0.5) {2};
\node at (-0.5,3-0.5) {$d$};
\node at (4-0.5,-0.5) {3};
\node at (-0.5,4-0.5) {\small{$d+1$}};
\node at (5-0.5,-0.5) {\dots};
\node at (-0.5,5-0.5) {\small{$d+2$}};
\node at (6-0.5,-0.5) {};

\node at (1.5,4.5) {\small{$[\alpha_1]$}};
\node (B) at (2.5,3.5) {\small{$[\alpha_2]$}};
\node at (3.5,2.5) {\small{$[\alpha_3]$}};
\node (D) at (4.5,1.5) {\small{$[\alpha_4]$}};

\node (A) at (0.5,4.5) {\small{$[f_0]$}};
\node at (1.5,3.5) {$[f_1]$};
\node (C) at (2.5,2.5) {$[f_2]$};
\node at (3.5,1.5) {$[f_4]$};

\draw[->, red, thick, bend left] (A) to node[midway, above, sloped, red] {} (B);
\draw[->, blue, thick, bend left] (C) to node[midway, above, sloped, red] {} (D);

\end{tikzpicture}
\caption{Schematic illustration of differentials in the compatibility AHSS.}
\label{subfig:Case1}

\end{figure}

The $E_2$-page of the compatibility AHSS is shown in Fig.~\ref{subfig:Case1}. Note that $E_2^{0,d+2} \cong h^{d+1}(\Lambda \times BH_0)$ classifies the $H_0$-families of interest; in particular, this includes all the $d$-dimensional pumps. To see this we use a general result 
\begin{equation}\label{eq:pump-decomposition}
    h^{n}(\Lambda \times BH_0) = h^{n}(BH_0) \times \tilde{h}^{n}_{H_0}(\Lambda).
\end{equation} 
The first factor denotes \textit{constant} families or ordinary $H_0$-invertible states, classified by $h^{d+1}(BH_0)$. The second denotes the families which become trivial when we set the order parameter to a constant value; these are precisely the pumps, and the group $\tilde{h}^n_{H_0}(\Lambda)$ is what we wrote as $\{\text{$H_0$-pumps over $\Lambda$}\}_{d}$ in Sec.~\ref{subsec:setup}.\footnote{Note that given a generalized cohomology $h^{\bullet}$, we can construct another generalized cohomology theory $h^{\bullet}_{H_0}$ such that $h^{\bullet}(X \times BH_0) = h^{\bullet}_{H_0}(X)$. The difference between these theories is in the definition of reduced cohomology: $\tilde{h}^{\bullet}(X \times BH_0)$ classifies the phases that are trivial when we forget $H_0$ and choose a constant value of $X$ (therefore it includes $H_0$-SPT phases), whereas $\tilde{h}^{\bullet}_{H_0}(X)$ classifies the phases that are trivial when we set $X$ to a constant value but do \textit{not} forget $H_0$. These are precisely the pumps.} We will denote a general term of $h^{d+1}(\Lambda \times BH_0)$ by $[f_0] = ([c],[p])$, where $[c] \in h^{d+1}(BH_0), [p] \in \tilde{h}^{d+1}_{H_0}(\Lambda)$. We will denote a general term of $E_2^{p,d+1-p}$ by $[f_p]$. Note that the decorations $f_0,f_1, \dots , f_{d+1}$ in the compatibility AHSS converge to $H$-invertible states classified by $h^{d+1}(BH)$.

Furthermore, each $G$- anomaly appears on the $p+q=d+2$ diagonal of the compatibility AHSS. 
Using standard rules for decomposing coefficients in cohomology, we have for each $k$
\begin{equation}
    H^k(BG^*,h^{d+2-k}(\Lambda)) \cong H^k(BG^*,h^{d+2-k}(BH_0)) \times H^k(BG^*,\tilde{h}^{d+2-k}_{H_0}(\Lambda)).
\end{equation}

We can see that the factors $H^k(BG^*,h^{d+2-k}(BH_0))$ are precisely the $E_2$-page terms in the $G$-spectral sequence defined above. As a result, a given anomaly $[\alpha] \in h^{d+2}(BG)$ can always be decomposed into classes $[\alpha_k] \in H^k(BG^*,h^{d+2-k}(BH_0))$, which will also appear in the compatibiliy AHSS.

We will generally choose $[\alpha_k]$ so as to correspond to some well-defined $G$ anomaly which is moreover trivial in $H$ (including the case with trivial anomaly). This implies that $[\alpha]$ must be \textit{trivialized} in the compatibility AHSS: there must be an incoming differential connecting it to a decoration on the $p+q=d+1$ diagonal.  (If it were instead removed by an outgoing differential, the putative $G$-anomaly would be obstructed in $H$ rather than trivial in $H$.) From this we see that $[\alpha_0]$ and $[\alpha_1]$ must be trivial if we want the anomaly to be compatible with a $d$-dimensional family. $[\alpha_0]$ denotes an $H_0$-anomaly, which must necessarily vanish, while $[\alpha_1]$ represents a decoration where a $G^*$ domain wall carries a $d$-dimensional $H_0$ family. It cannot be trivialized by any incoming differential from the $d$-dimensional decorations. The only possibilities are that it survives as an $H$-anomaly, or that it is obstructed by an outgoing differential, meaning that the decoration breaks $H$ symmetry. Either way, such a decoration must be ruled out as a choice for $[\alpha]$.

The differentials in the compatibility AHSS completely encode the compatibility condition we need. To see this, let us consider four basic situations. The general case can be understood by combining these base cases. Note that the extension problems in spectral sequences make it difficult to prove the stacking properties of the compatibility relation directly from the compatibility AHSS. However, since they can be inferred through separate arguments that do not require spectral sequences (see Appendix~\ref{app:Homotopy}), we will take them as given in the discussion below.

\begin{enumerate} 
\item Case 1: First consider the case with trivial $G$-anomaly, $[\alpha_k] = [0]$ for all $k$. In this case, the only compatible pumps are the ones which survive without getting obstructed by any differentials. More generally, the full set of families compatible with the trivial anomaly (including constant families) are the elements which survive on the entire $p+q=d+1$ diagonal. These elements are in one-to-one correspondence with $H$-invertible states, in agreement with the general result of Ref.~\cite{else2021goldstone}. Case 1 already shows that two distinct pumps can be compatible with the same anomaly, if their difference under stacking is compatible with the trivial anomaly.

It is possible that we fix a pump $[p]$ and find that $d_k [p] = [\alpha_k]$ for some $[\alpha_k]$ which is nontrivial on the $E_2$-page. However, upon running the $G$-spectral sequence, we may find that $[\alpha_k]$ is in the \textit{trivial} $G$-anomaly class. In this case we would still say that $[p]$ is compatible with a trivial anomaly.
\item Case 2: Now assume the anomaly is non-trivial and is completely captured by a single term $[\alpha_k]$ on the compatibility AHSS. Suppose there is a differential $d_k [p] = [\alpha_k]$ for some pump $[p]$. For example, see the red differential in Fig.~\ref{subfig:Case1}. In this case, we write $[p] \tl [\alpha]$. This is the most direct relation between a non-trivial anomaly and a non-trivial pump.

Whenever a pump satisfies $d_k[p] = [\alpha_k]$, we also have $d_k([p] + [c]) = [\alpha_k]$, where $[c]$ denotes any constant $H_0$-symmetric family whose differentials are all zero. Here, the same anomaly is compatible with multiple pumps that only differ by an $H_0$-invertible state.

\item Case 3: Suppose the anomaly $[\alpha]$ is captured by a single term, say $[\alpha] = [\alpha_4]$, on the compatibility AHSS. But now, assume it is trivialized by a differential $d_2$, say $d_2[f_2] = [\alpha_4]$ for some decoration $[f_2]$. See the blue arrow in Fig.~\ref{subfig:Case1}. In this case, we argue that the trivial pump $[0]$ satisfies $[0] \tl [\alpha]$. This is because the entries $f_1, f_2, \dots $ of the $E^2$ page correspond to decorating families on $G^*$ domain walls, while the definition of pump is completely agnostic about the properties of $G^*$ domain walls. 
Combining this with the usual consistency conditions, we then obtain that every pump which is compatible with the trivial anomaly is also compatible with $[\alpha]$.

The observations in Cases 1 and 3 imply that we generally do not have a well-defined map from the group of invertible pumps to the group of anomalies, and the above compatibility relation is the best we can define.
 
\item Case 4: In the three cases above we fixed some anomaly and asked which pumps were compatible with it. If we instead fix a pump and investigate the compatible anomalies, there is one additional case to consider: it could be that there is a differential $d_k$ such that $d_k [p]$ is nonzero but also not a well-defined $G$-anomaly term. For example, it could be order parameter dependent. In this case, we would conclude that the given pump is not compatible with \textit{any} $G$-anomaly (not even the trivial anomaly), since it can only exist on the boundary of a higher-dimensional family.

\subsection{Spectral sequence computations for Sec.~\ref{sec;Apps}}\label{app:Apps}

In this section we apply the compatibility AHSS to the examples discussed earlier in Sec.~\ref{sec;Apps}. These examples illustrate all the cases enumerated above. In particular, the Chern number pump is an instance of Case 4: it is only compatible with another family in $d=2$, but not with a $G$-anomaly. 

\subsubsection{Assumptions}

Before proceeding, we should clarify that it is very difficult to rigorously derive the form of each differential for arbitrary $G,H,\Lambda$. We generally have to resort to physical arguments to infer these differentials, or we might be fortunate that they clearly vanish by inspection in examples of interest. Often we need to pick a convenient formulation of the AHSS where sufficiently many differentials can be seen to vanish. 

One specific assumption that we will make in the context of invertible fermionic states (see the example of the QSH insulator, sec.~\ref{app:QSH}) is the following. The differentials for the $G$-spectral sequence for invertible fermionic states are known in dimensions $d \le 3$ from the results of Refs.~\cite{Wang2020fSPT,barkeshli2021invertible,aasen2021characterization}. We will assume that certain differentials in the compatibility AHSS involving the constant families have the same form as in the $G$ spectral sequence. (The differentials involving the pumps need to be determined separately.) While there is no physical reason to doubt this assumption, we do not have a mathematical proof of it. We can check that this assumption gives results that are consistent with the independent physical arguments in Sec.~\ref{sec:Apps}.

\subsubsection{Notation}

We first define a number of generating classes for low-dimensional cohomology groups that appear in our examples. Let $[\Phi_n]$ generate $H^n(S^n,\Z) = \Z$; note that $[\Phi_n^2]$ is trivial in cohomology. Let $[\nu_2]$ generate $H^2(B\SO(2),\Z) = \Z$; note that $H^{2k}(B\SO(2),\Z)$ is generated by $[\nu_2^k]$. Let $[w_2]$ generate $H^2(B\SO(3),\Z_2) = \Z_2$, so that $[w_3]$ (which is the image of $[w_2]$ under a Bockstein map) generates $H^3(B\SO(3),\Z) = \Z_2$. Let $[p_1]$ be the Pontryagin class, which generates $H^4(B\SO(3),\Z) = \Z$. Let $[u_4]$ generate $H^4(B\SU(2),\Z) = \Z$. For $\Z_2^T$ time-reversal symmetry, we define $[\Tt_1], [\Tt_2]$ as the generators of $H^1(B\Z_2^T,\Z^{\text{or}})$ and $H^2(B\Z_2^T,\Z)$ respectively. Note that the action on coefficients is non-trivial in the first case.

\subsubsection{Berry-Chern number families in $d=0$}
In this case, we consider bosonic systems with $G = \SO(3), H = \SO(2), \Lambda = S^2$ and $H_0$ trivial. We consider the fibration
\begin{equation}
    S^2 \rightarrow B\SO(2) \rightarrow B\SO(3),
\end{equation}
and the associated compatibility AHSS
\begin{equation}
    H^p(B\SO(3),H^q(S^2,\Z)) \implies H^{p+q}(B\SO(2),\Z).
\end{equation}

\begin{center}
    \begin{tikzpicture}[>=stealth]
  \matrix (m) [matrix of math nodes,
    nodes in empty cells,nodes={minimum width=5ex,
    minimum height=5ex,outer sep=-5pt},
    column sep=1ex,row sep=1ex]{
       
          4   &  0 & 0 & 0 & 0 & 0\\
           3   & 0 &  0 &  0 & 0 & 0 \\
           2     &  ^{\Phi_2}\Z    & 0    &   0   & ^{\Phi_2 w_3}\Z_2 & ^{\Phi_2 p_1}\Z \\
          1     &  0 &  0  & 0  & 0 & 0 \\
          0     &  \Z  & 0 &  0  & ^{w_3}\Z_2 & ^{p_1}\Z \\
    \qquad\strut &   0  &  1  &  2  & 3 & 4 \strut \\};
\draw[thick] (m-1-1.east) -- (m-6-1.east) ;
\draw[thick] (m-6-1.north) -- (m-6-6.north) ;
\end{tikzpicture}
\end{center}
There are only two nonzero rows on the compatibility AHSS. Note that $[w_3] \in H^3(B\SO(3),\Z) \cong \Z_2$ corresponds to the classification of 1d SPT states (or 0d anomalies) with $\SO(3)$ symmetry. The $p+q=3$ diagonal of the AHSS converges to the classification $H$-symmetric invertible states in $d=1$. But since $H = \U(1)$, this should be trivial. Therefore we are forced to have a nonzero differential $d_3$ that satisfies $d_3[\Phi_2] = [w_3]$. 

This map states that the generator of the Berry-Chern number families is compatible with the nontrivial anomaly class. Futhermore, the kernel of this map is a $\Z$ factor generated by $2[\Phi_2]$; this factor survives to generate the classification of $\SO(2)$ SPT states in $d=0$. As a result, we can express the full compatibility relation as 
\begin{equation}
    [C] \tl [\alpha] \Leftrightarrow C = \alpha \mod 2 .
\end{equation}

\subsubsection{Thouless pump}\label{app:Thouless}

We take $G = \U(1)_a \times \U(1)_b, H = \U(1)_b = H_0, \Lambda = S^1$. We consider the fibration 
\begin{align}
   & \Lambda \rightarrow B(H/H_0) \rightarrow B(G/H_0) \nonumber \\
   \Leftrightarrow \quad &S^1 \rightarrow B\Z_1 \rightarrow B \U(1)_a
\end{align}
with the compatibility AHSS defined by
\begin{align}\label{eq:AHSS-Thouless}
    H^p(B\U(1)_a, h^q_{\U(1)_b}(S^1)) \implies h^{p+q}_{H_0}(\text{pt}) .
\end{align}
The generalized cohomology theory $h^{\bullet}_{\U(1)_b}$ is defined so that $h^{d+1}_{\U(1)_b}(\text{pt})$ classifies the $\U(1)_b$-symmetric invertible states in $d$ dimensions. Although the fermionic symmetry $G_f$ in this example is not of the form $G \times \Z_2^f$, it is still of the form described in Footnote~\ref{footnote:more_general} from Section \ref{sec:MainRes-TrivAnom}, which implies that we can still use an AHSS based on Eq.~\eqref{eq:AHSS-Thouless}. $h^{d+1}_{H_0}(\text{pt})$ equals $\Z$ in $d=0$ and $d=2$, corresponding to $\U(1)_b$ charges and the integer quantum Hall states. This applies to both bosonic and fermionic systems.\footnote{There are really two $\Z$ factors in $d=2$ which correspond to the IQH states and multiples of the $E_8$ invertible state. We will ignore the second factor here since it does not affect our discussion of the Thouless pump.} The nonzero $E_2$-page elements are $E_2^{2m,n} := H^{2m}(B\U(1)_a, h^n_{\U(1)_b}(S^1)) \cong \Z$. Therefore the $E_2$-page is filled with $\Z$'s and 0's on alternating columns.

We can express the $E_2$-page generators in terms of their effective actions as shown below for some relevant classes (overall factors of $2\pi$ are not shown); this is done by pulling back the relevant $\U(1)$ and $S^1$ cohomology classes using the gauge fields $a,b$ and the order parameter field $\phi$. In the compatibility AHSS, we write the topological terms with $\Z$ coefficients instead of the usual $\U(1)$ coefficients, which amounts to writing $d\mathcal{L}$ instead of $\mathcal{L}$:

\begin{center}
    \begin{tikzpicture}[>=stealth]
  \matrix (m) [matrix of math nodes,
    nodes in empty cells,nodes={minimum width=5ex,
    minimum height=5ex,outer sep=-5pt},
    column sep=1ex,row sep=1ex]{
       4   &  \Z    & 0    &  \Z   & 0& \Z\\
          3   &  ^{db db}\Z    & 0    &  \Z   & 0& \Z\\
           2  &   ^{d\phi db}\Z    & 0    &  \Z   & 0& \Z\\
           1     &  ^{db}\Z    & 0    &  ^{da db}\Z   & 0& \Z\\
          0      &  ^{d\phi}\Z    & 0    &  ^{d\phi da}\Z   & 0& \Z\\
          -1      &  \Z    & 0    &  ^{da}\Z   & 0& ^{da da}\Z\\
    \qquad\strut &   0  &  1  &  2  & 3 & 4 \strut \\};
\draw[thick] (m-1-1.east) -- (m-7-1.east) ;
\draw[thick] (m-7-1.north) -- (m-7-6.north) ;

\draw[->, red, thick, bend left] (m-3-2.east) to node[midway, above, sloped, red] {} (m-4-4.north);
\end{tikzpicture}
\end{center}

Elements on the $p+q=d+1$ diagonal represent decorations in $d$ space dimensions. The only terms which survive in the spectral sequence should represent $H$-SPTs, which are pure functionals of $b$ and correspond to the terms $E_2^{0,2k+1}, k = 0,1, \dots$. This means that the terms depending on $\phi, a$ must be cancelled by differentials which map generators to generators. In particular $d_2(d\phi) = da$; all other differentials on the $E_2$-page follow from the Leibniz rule for differentials of products of cocycles. The physical interpretation of this is that an elementary order parameter vortex inherits the quantum numbers of a $2\pi$ flux of $\U(1)_a$. The differential connecting the Thouless pump response to its anomaly term is shown in red; it is an isomorphism $\Z \rightarrow \Z$. From this differential we conclude that if a given pump $[p]$ has coefficient $m$, it is compatible with an anomaly term $[\alpha] = (n_1,n_2,n_3)$ when $n_1 = n_3 = 0$ and $n_2 = m$.

Next, note that $n_3$, the coefficient of the $b db$ anomaly term, should always be set to zero to have a well-defined pump, since it is also a non-trivial $H$-anomaly. Finally, the $a da$ anomaly term is compatible with a trivial $\U(1)_b$ charge pump, since it does not depend on the $\U(1)_b$ symmetry at all. This means that $[0] \tl (n_1,0,0)$. Combining these results, we get Eq.~\eqref{eq:Thouless-CR} in the main text.

\subsubsection{Computations for quantum spin Hall insulator}\label{app:QSH}
In this case, we have $G = \frac{\U(1)^f \rtimes \Z_4^{Tf}}{\Z_2^f}$, with $G_b = \U(1) \rtimes \Z_2^T$. We have $H = \Z_4^{Tf}, H_b = \Z_2^T$, with $\Lambda = S^1$. Finally, since we subsume the fermion parity symmetry into the definition of the AHSS, $H_0 = \Z_1$.

The spectral sequence written below does not fit within the assumptions of Sec.~\ref{sec:MainTechRes}, because the fermionic symmetry $G_f$ is not of the form $G \times \Z_2^f$ (nor of the form discussed in Footnote \ref{footnote:more_general} in Section \ref{sec:MainRes-TrivAnom}). However, we expect that it should exist on physical grounds, since there is a decorated domain wall interpretation for it. We assume that there exists a compatibility AHSS associated to the fibration $ S^1 \rightarrow B \Z_2^{T} \rightarrow B (\U(1) \rtimes \Z_2^T)$, with the form
\begin{equation}
   H^p(B(\U(1) \rtimes \Z_2^T), h^q(S^1)) \implies h^{p+q}_{\Z_4^{Tf}}(\text{pt}).
\end{equation} 
$h^{p+q}_{\Z_4^{Tf}}(\text{pt})$ corresponds to the classification of invertible fermionic states in spatial dimension $p+q-1$ with symmetry $\Z_4^{Tf}$. The notation for the left-hand side is a heuristic notation which represents the appropriate decorated domain wall construction for this example, including fermion parity twists. 

The relevant elements on the $E_2$-page of the compatibility AHSS have the form

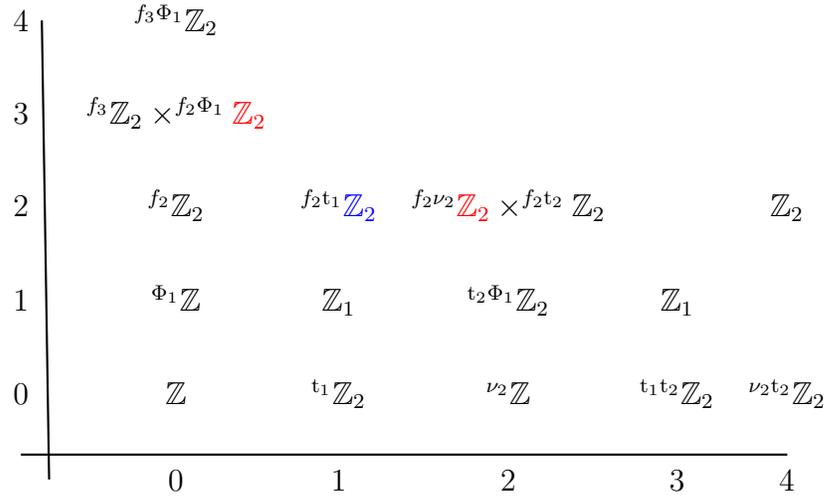
\begin{center}
    \begin{tikzpicture}[>=stealth]
  \matrix (m) [matrix of math nodes,
    nodes in empty cells,nodes={minimum width=5ex,
    minimum height=5ex,outer sep=-5pt},
    column sep=1ex,row sep=1ex]{
          4   &   ^{f_3\Phi_1}\Z_2  &     &     &  & \\
           3   &  ^{f_3}\Z_2\times^{f_2\Phi_1} \textcolor{red}{\Z_2}   &     &     &  &  \\
           2     &  ^{f_2}\Z_2    &  ^{f_2 \Tt_1}\textcolor{blue}{\Z_2}   &  ^{f_2 \nu_2}\textcolor{red}{\Z_2} \times ^{f_2 \Tt_2}\Z_2   &  & \Z_2 \\
          1     &  ^{\Phi_1}\Z &  \Z_1  & ^{\Tt_2 \Phi_1}\Z_2  & \Z_1 &  \\
          0     &  \Z  & ^{\Tt_1}\Z_2 &  ^{\nu_2}\Z  & ^{\Tt_1\Tt_2}\Z_2  & ^{\nu_2 \Tt_2}\Z_2\\
    \qquad\strut &   0  &  1  &  2  & 3 & 4 \strut \\};
\draw[thick] (m-1-1.east) -- (m-6-1.east) ;
\draw[thick] (m-6-1.north) -- (m-6-6.north) ;
\end{tikzpicture}
\end{center}

Some comments on notation: We denote by $\Tt_1, \Tt_2, \nu_2$ the generators of $\HH^1(\Z_2^T, \Z^{\text{or}})$ (with the orientation reversing action on coefficients), $\HH^2(\Z_2^T, \Z)$ and $\HH^2(\U(1),\Z)$ respectively (the latter two are defined for trivial action on coefficients). We will abuse notation slightly and also denote the generators of $\HH^1(\Z_2^T,\Z_2)$ and $\HH^2(\Z_2^T,\Z_2)$ by $[t_1]$ and $[t_2]$. The elements $[f_2], [f_3]$ correspond to the nontrivial invertible fermionic states in $d=0$ and $d=1$ (a complex fermion and a Kitaev chain respectively). We know that the spectral sequence should return the classification of invertible states in Class DIII with symmetry $\Z_4^{Tf}$, which is $\Z_1$ in $d=0$ and $\Z_2$ in $d=1$. Therefore we should expect differentials to trivialize several terms; the elements that survive are shown in blue. 

First, a $\pi$ flux of $\U(1)^f$ is equivalent in the group sense to a $2\pi$ flux of the bosonic $\U(1)$ symmetry appearing on the AHSS. But a $\pi$ flux also creates an $S^1$ order parameter vortex in one lower dimension; this is one way to see that $d_2 [\Phi_1] = [\nu_2]$. Next, the Kitaev chain corresponding to $[f_3]$ does not survive as a Class DIII state, because the existence of edge Majorana zero modes is incompatible with the condition $T^2 = (-1)^F$. Therefore $d_2[f_3] \neq 0$. The meaning of this differential going into $E_2^{2,2}$ is that certain decotations for the 2d state are equivalent under a relabelling of fermion parity fluxes by fermions. It is known that $E_2^{2,2}$ describes a 2d decoration which sets the quantum numbers of fermion parity fluxes (the corresponding cochain is referred to as $n_2 \in C^2(G_b,\Z_2)$ in the classification of 2d invertible fermion states \cite{Wang2020fSPT,barkeshli2021invertible}). We know that a fermion carries both 1/2 charge under the bosonic $\U(1)$ and a Kramers degeneracy, so the relabelling should take $n_2 \rightarrow n_2 + \nu_2 + \Tt_2$. The entire relation can be expressed in our notation as $d_2 [f_3] = [f_2(\nu_2 + \Tt_2)]$. (This is where we use the assumption that the differential in the compatibility AHSS involving the Kitaev chain is identical to the known differential in the $G$ spectral sequence.) Finally, the differential $d_2$ does not affect the complex fermion state.

These relations are enough to understand the fate of the nontrivial fermion pump family, which is an element of $E_2^{0,3}$ represented by the class $[f_2 \Phi]$. We have $d_2 [f_2 \Phi] = [f_2] d_2 [\Phi] = [f_2 \nu_2]$. But this corresponds to the anomaly of the QSH insulator. The domain and target of the relevant differential is shown in red on the AHSS. Therefore we have shown that the QSH boundary anomaly is compatible with a fermion pump: $[p] \tl [\alpha]$. The same differential also shows that $[p] \cancel{\tl} [0]$ and $[0] \cancel{\tl} [\alpha]$. 

Finally, note that the constant families are generated by the term $E_2^{2,1}$, which corresponds to a decoration for the Class DIII state in 1d. In this case, the $H$-invertible states do \textit{not} lie in $E_2^{0,d+2}$, meaning that they do not correspond to $H_0$-invertible states.

\subsubsection{Analysis for quantum Hall ferromagnet}\label{app:QHFM}
Recall that the parameters for this example are $G = \U(2), H = \U(1) \times \U(1)$ and $\Lambda = S^2$. In this case, $G^* = \SU(2), H^* = \U(1)$ and $H_0 = \U(1)_c$ (the subgroup corresponding to charge conservation). We construct the compatibility AHSS using the fibration
\begin{equation}
    S^2 \rightarrow B \U(1) \rightarrow B\SU(2), 
\end{equation}
in which we have quotiented out by $\U(1)_c$. It is most convenient to define the compatibility AHSS as follows:
\begin{equation}
    H^p(B\SU(2), h^q_{\U(1)_c}(S^2)) \implies h^{p+q}_{\U(1)_c}(B\U(1)).
\end{equation}
Defining the compatibility AHSS in terms of the classification of $\U(1)_c$-invertible states, which is $h^{\bullet}_{\U(1)_c}$, ensures that several unimportant details relating to Kitaev chains and other obstructed states do not appear in the calculation below.

We use the facts that $h^n_{\U(1)_c}(S^2) = h^n_{\U(1)_c}(\text{pt}) \times h^{n-2}_{\U(1)_c}(\text{pt})$, with $h^n_{\U(1)_c}(\text{pt}) = \Z$ for $n=1$, and $\Z^2$ for $n=3$. Physically, $h^1_{\U(1)_c}$ classifies the group of $\U(1)$ charges, i.e. invertible states in $d=0$, while $h^3_{\U(1)_c}$ classifies $d=2$ invertible states with $\U(1)_c$ symmetry. There are two $\Z$ factors, generated by the IQH state with chiral central charge 1 and the bosonic IQH state, which is non-chiral. Now, the AHSS simplifies as follows:
\begin{center}
    \begin{tikzpicture}[>=stealth]
  \matrix (m) [matrix of math nodes,
    nodes in empty cells,nodes={minimum width=5ex,
    minimum height=5ex,outer sep=-5pt},
    column sep=1ex,row sep=1ex]{
       
          4   & ^{\Phi_2 \nu_2}\Z \times \Z^2    &  0   & 0    & 0 & \Z^3\\
           3   &   0   &  0   & 0    & 0 & 0 \\
           2     &   ^{\Phi_2}\Z \times ^{\nu_2}\Z    & 0    &   0  & 0 & \Z\\
          1     &  0 &  0  & 0  & 0 & 0 \\
          0     &  \Z  & 0 &  0  & 0 & ^{u_4}\Z \\
    \qquad\strut &   0  &  1  &  2  & 3 & 4 \strut \\};
\draw[thick] (m-1-1.east) -- (m-6-1.east) ;
\draw[thick] (m-6-1.north) -- (m-6-6.north) ;
\end{tikzpicture}
\end{center}
Here we have defined the generating classes $[\Phi_2] \in H^2(S^2,\Z), [\nu_2] \in H^2(B\U(1)_c,\Z), [u_4] \in H^4(B\SU(2),\Z)$. For dimensional reasons, there can be no nontrivial differentials outgoing from $[\Phi_2]$. As a result, the pump described by $[\Phi_2 \nu_2]$ carries the topological invariants of an $H^*$-invertible state.

\subsubsection{$\Lambda = S^3$ and higher Chern number}\label{app:S3}
Since $S^3 \cong \SU(2)$ as a topological space, we can define a fibration and its associated compatibility AHSS with $G = \SU(2), H = \Z_1$ and $\Lambda = S^3$:
\begin{align}
    & S^3 \rightarrow B\Z_1 \rightarrow B\SU(2) \\
    & H^p(B\SU(2),H^q(S^3,\Z)) \implies H^{p+q}(B\Z_1,\Z).
\end{align}
The $E_2$-page of the compatibility AHSS is constructed by using the result that $\HH^n(\SU(2),\Z) = \Z$ for $n=0,4$ and is trivial for $n=1,2,3$:
\begin{center}
    \begin{tikzpicture}[>=stealth]
  \matrix (m) [matrix of math nodes,
    nodes in empty cells,nodes={minimum width=5ex,
    minimum height=5ex,outer sep=-5pt},
    column sep=1ex,row sep=1ex]{
       \vdots   &      &     &     & & \\
          3   &  ^{\Phi_3}\Z    & 0    &     & &\\
           2     &  0    &   0  &  0   & & \\
          1     &  0 &  0 & 0 & 0 & \\
          0     &  \Z  & 0 &  0  & 0 & ^{u_4}\Z \\
    \qquad\strut &   0  &  1  &  2  & 3 & 4 \strut \\};
\draw[thick] (m-1-1.east) -- (m-6-1.east) ;
\draw[thick] (m-6-1.north) -- (m-6-6.north) ;
\end{tikzpicture}
\end{center}
The above AHSS must converge to the classification of bosonic SPTs with no symmetry, which is trivial in all dimensions by definition. Thus the pump invariant, which is generated by $[\Phi_3] \in H^3(S^3,\Z)$, must be killed by a differential whose image lies on the $p+q=4$ diagonal. But the only nonzero term on this line is a $\Z$ factor generated by $[u_4] \in H^4(B\SU(2),\Z)$. Therefore we must have $d_4[\Phi_3] = [u_4]$. In fact this map is an isomporphism, therefore it can be inverted to give $\varphi: \Z \rightarrow \Z$, which takes $[u_4] \rightarrow [\Phi_3]$.

\subsubsection{$\Lambda = S^2 \times S^1$ and Chern number pump}\label{app:Chern_pump}
Here we use the fibration $ \Lambda \rightarrow BH \rightarrow BG$, with $G = \U(1) \times \SO(3)$ and $H = \SO(2)$. We consider the compatibility AHSS defined by
\begin{equation}
    H^p(BG,H^q(\Lambda,\Z)) \implies H^{p+q}(BH,\Z).
\end{equation}
We will adopt the same conventions as in previous computations.
\begin{center}
    \begin{tikzpicture}[>=stealth]
  \matrix (m) [matrix of math nodes,
    nodes in empty cells,nodes={minimum width=5ex,
    minimum height=5ex,outer sep=-5pt},
    column sep=1ex,row sep=1ex]{
       
          4   & 0    &  0   & 0    & 0 & 0\\
           3   &   ^{\Phi_1\Phi_2}\Z   &  0   & \Z    & \Z_2 & \Z^2 \\
           2     &   ^{\Phi_2}\Z    & 0    &   ^{\Phi_2 \nu_2}\Z  & ^{\Phi_2 w_3}\Z_2 & \Z^2\\
          1     &  ^{\Phi_1}\Z &  0  & ^{\Phi_2 \nu_2}\Z  & ^{\Phi_2 w_3}\Z_2 & \Z^2 \\
          0     &  \Z  & 0 &  ^{\nu_2}\Z  & ^{w_3}\Z_2 & ^{p_1}\Z \times ^{\nu_2^2}\Z \\
    \qquad\strut &   0  &  1  &  2  & 3 & 4 \strut \\};
\draw[thick] (m-1-1.east) -- (m-6-1.east) ;
\draw[thick] (m-6-1.north) -- (m-6-6.north) ;
\end{tikzpicture}
\end{center}
The main differentials are $d_2 [\phi_1] = [\nu_2]$ and $d_3 [\Phi_2] = [w_3] \mod 2$, which can be read off from our previous calculations on the Berry-Chern number family and the Thouless pump. This implies that $d_2 [\Phi_1 \Phi_2] = [\Phi_2 \nu_2]$, which is a $d=2$ family over $S^2$ with $\U(1)$ symmetry.  

The lowest dimension in which there is a non-trivial mixed anomaly of $G$ is $d=3$; this anomaly is $\Z_2$ classified and generated by the class $[\nu_2 w_3]$. By breaking \textit{either} the $\U(1)$ \textit{or} the $\SO(3)$ symmetry, we can see that this anomaly is compatible with either a `Haldane pump', in which an $S^1$ order parameter defect traps a Haldane chain in $d=1$, or the charge pump over $S^2$ mentioned in the main text.

\end{enumerate}

\section{The case of non-invertible families}\label{sec:MainRes-noninv}

In Sec.~\ref{sec:GC-Classif} we saw that the classification of invertible families can be obtained from that of invertible states by replacing $BH_0$ for some $H_0$ with the space $\Lambda \times BH_0$ associated to the family. The classification of topologically ordered families can be handled analogously, as has been discussed recently \cite{Hsin_2023moduli,aasen2021characterization}. For concreteness, we consider bosonic SET phases in $d=2$. In this case the classification of $G$ symmetry-enriched topological states is captured by three levels of data \cite{barkeshli2019}: 
\begin{enumerate}
    \item A homomorphism $\rho: G \rightarrow \text{Aut}(\mathcal{C})$, where $\text{Aut}(\mathcal{C})$ denotes the group of automorphisms of the anyon theory $\mathcal{C}$; this encodes how the symmetry permutes anyons.
    \item A set of $F,R,U,\eta$ symbols for the anyons. This fixes the symmetry fractionalization class (or an obstruction to defining such a class). The symmetry fractionalization is classified by $\HH^2_{\rho}(G,\mathcal{A})$ where $\mathcal{A}$ denotes the group of Abelian anyons, while the obstruction to it is classified by an element of $\HH^3_{\rho}(G,\mathcal{A})$. 
    \item Note that there is a 't Hooft anomaly for $G$ which is fixed by (1) and (2). If this anomaly is trivial we can further consistently define a set of $F, R, U, \eta$ symbols for the $G$ symmetry defects, which fix the overall topological response. Finally, given a consistent set of data for an SET phase, we can stack a $d=2$ bosonic SPT classified by $\HH^4(G,\mathrm{U}(1)) \cong H^4(BG, \mathbb{Z})$ to obtain another consistent set of data which describes a possibly distinct SET.
\end{enumerate}

Ref.~\cite{Hsin_2023moduli} studied the analogous classification of topologically ordered families, and its main result can be stated as follows. Given a parameter space $\Lambda$ and a symmetry $H_0$ preserved by the entire family, the classification consists of the following data:
\begin{enumerate}
    \item A map $\gamma : \Lambda \times BH_0 \rightarrow B \text{Aut}(\mathcal{C})$. Note that this induces a homomorphism $\pi_1(\Lambda \times BH_0) \to \pi_1(B \text{Aut}(\mathcal{C})) \cong \text{Aut}(\mathcal{C})$, which specifies the anyon permutation activated by dragging an anyon across a codimension-1 $H_0$ defect worldsheet or a codimension-1 order parameter defect. 
    \item Restricting the image of the above homomorphism to $\text{Aut} (\mathcal{A})$ defines an action on $\mathcal{A}$, which we denote as $\gamma_{\mathcal{A}}$. We can now define a symmetry fractionalization class, given by an element of $H^2_{\gamma_{\mathcal{A}}}(\Lambda \times BH_0, \mathcal{A})$. Here $\gamma_{\mathcal{A}}$ defines a ``system of local coefficients'' on $\Lambda \times BH_0$, with respect to which the cohomology is defined.
    \item There is a 't Hooft anomaly fixed by the symmetry fractionalization data, which if trivial leads to equivalence classes of defect data. These form a torsor over the group of bosonic invertible families, which is $H^4(\Lambda \times BH_0, \Z)$. 
\end{enumerate}

As discussed in Sec.~\ref{subsec:noninvertible}, the compatibility relation for non-invertible families with anomaly-free $H$ can be obtained using the results for invertible families. But even when the $H$ symmetry fractionalization is anomalous, we can derive some conclusions about the symmetry fractionalization of the family using the tools described in prior sections. An action of $H$ on $\mathcal{C}$ is specified by a map $\rho: \pi_1(BH) \rightarrow \text{Aut}(\mathcal{C})$; note that $\pi_1(BH) \cong \pi_0(H)$. 
When we pull back $\rho$ using the projection $i: \Lambda \times BH_0 \rightarrow BH$, we obtain a map $i^* \rho : \pi_1(\Lambda \times BH_0) \rightarrow \text{Aut}(\mathcal{C})$. This map defines an induced action of $\Lambda \times BH_0$ on $\mathcal{\mathcal{A}}$, which can be taken as the definition of $\gamma_{\mathcal{A}}$. Next, given an $H$ fractionalization class specified by an element of $H^2_{\rho}(BH,\mathcal{A})$, we can again pull back the map $i$ to obtain a map $i^*: H^2_{\rho}(BH,\mathcal{A}) \rightarrow H^2_{\gamma_{\mathcal{A}
}}(\Lambda \times BH_0, \mathcal{A})$. This specifies a symmetry fractionalization class for the family which is consistent with $\gamma_{\mathcal{A}}$. These relations can be defined irrespective of whether the symmetry fractionalization in $H$ is anomalous or not.

\section{Discussion}\label{sec:Disc}

\subsection{Summary}
In this work we have shown how define compatibility relations between a $G$-anomaly and the Goldstone invariants (or equivalently the family invariants) that can be realized when $G$ is spontaneously broken to an anomaly-free subgroup $H$. This generalizes the result of Ref.~\cite{else2021goldstone}, which showed that the Goldstone invariants compatible with a trivial $G$ anomaly are fully described by SPT/SET invariants of the residual $H$ symmetry. An important point is that the compatibility relation is generally only a binary relation between families over $\Lambda = G/H$ and $G$-anomalies; it acquires additional structure (such as that of a group homomorphism) only in special cases, for example when $G = \hat{G} \times H$ (see Sec.~\ref{subsec:simple_anomaly}). We introduced a `compatibility spectral sequence' which can be used to compute the desired relation in the invertible case. Then we showed that the non-invertible case can be handled as a corollary if we assume that $H$ is anomaly-free, but made some partial progress even in the general case.

\subsection{Relation to homotopy long exact sequence}
Note that the existence of a fibration $\Lambda \rightarrow BH \rightarrow BG$ implies the homotopy long exact sequence
\begin{align}\label{eq:homotopyLES}
\dots \rightarrow & \pi_{k+1}(\Lambda) \rightarrow \pi_{k+1}(BH) \rightarrow \pi_{k+1}(BG) \nonumber \\
\rightarrow & \pi_{k}(\Lambda) \rightarrow \pi_{k}(BH) \rightarrow \pi_{k}(BG) \rightarrow \dots
\end{align}
The usual homotopy long exact sequence for the fibration $H \rightarrow G \rightarrow \Lambda$ can be recovered from this by replacing $\pi_k(BH)$ with $\pi_{k-1}(H)$, and similarly for $G$. Eq.~\eqref{eq:homotopyLES} has the interpretation that order parameter defects in codimension $k$ (classified by $\pi_k(\Lambda)$) are related either to $G$ gauge field defects in one lower dimension (classified by $\pi_{k+1}(BG)$) or to $H$ gauge field defects in the same dimension ($\pi_{k}(BH)$). A more detailed physical understanding is given in Ref.~\cite{mermin1979}. This is a statement about classical defect configurations; there is no `quantum topology' encoded in the homotopy exact sequence. But if we are given additional information about the family (e.g. that an order parameter defect binds some $H_0$-SPT), we can make further conclusions. For example, if the order parameter defect is related to a $G$ gauge field defect in one lower dimension, the $G$ gauge field defect must carry the boundary modes of the $H_0$-SPT; this indicates a $G$-anomaly, since such a defect can only be the termination of some higher-dimensional object. On the other hand, if the order parameter defect is related to an $H$ gauge field defect, the family invariant can be thought of as an $H$-SPT invariant. 

The homotopy exact sequence was used previously in Ref.~\cite{else2021goldstone}; it provides a quick and intuitive way to determine whether a family invariant is related to a $G$-anomaly or to an $H$-SPT, at least in simple examples. However, it is less general, since the generalized cohomology classification of invertible states/families is not in one-to-one correspondence with that of SPTs bound to to gauge field/homotopy defects. Therefore, in order to be more general, we prefer to use the language of generalized cohomology in this work.

\section{Acknowledgements}
We thank Zhen Bi, Tarun Grover, Po-Shen Hsin, Andrew Potter, Ryan Thorngren and Rui Wen for helpful discussions. Research at Perimeter Institute is supported in part by the Government of Canada through the Department of Innovation, Science and Economic Development and by the Province of Ontario through the Ministry of Colleges and Universities.
\appendix
\clearpage
\section{Derivation of Eq.~\eqref{eq:fibration}}\label{app:Proofs}

In order to prove Eq.~\eqref{eq:fibration} and apply the AHSS to it, we need to show that there exists a fibration $\pi: BH \rightarrow B(G/H_0)$  with fiber $\Lambda \times BH_0$. We will require the following results:

\textbf{Lemma 1}: \textit{Given a group $G$, a subgroup $H$, and a space $\Lambda \simeq G/H$ on which $G$ acts transitively, there is a fibration $\pi: BH \rightarrow BG$ with fiber $\Lambda = G/H$.} 

\textbf{Proof (reference)}: See Theorem 11.3 of Ref.~\cite{Mitchell2006NotesOP}. To sketch the idea, a standard result is that the homotopy quotient $\Lambda//G$ defined in the main text fits into the following fibration:
\begin{equation}
    \Lambda \rightarrow \Lambda//G \rightarrow BG.
\end{equation}
But since $\Lambda//G \simeq BH$ (see Ref.~\cite{else2021goldstone}, or Corollary 3.4 of Ref.~\cite{Mitchell2006NotesOP}), we obtain the desired result.

\textbf{Lemma 2}: \textit{Given a group $G$ and a normal subgroup $H_0$, there is a fibration $\tau: BG \rightarrow B(G/H_0)$ with fiber $BH_0$.}

\textbf{Proof (reference)}: See Theorem 11.4 of \cite{Mitchell2006NotesOP}.

We apply Lemmas 1 and 2 and use the fact that the composition of two fibrations is also a fibration. This means that $\tau \circ \pi: BH \rightarrow B(G/H_0)$ is a fibration. It remains to show that the fiber $(\tau \circ \pi)^{-1}(b), b \in B(G/H_0)$, equals $\Lambda \times BH_0$. To see this, it is convenient to write $B(G/H_0) \cong E(G/H_0)/(G/H_0)$ as usual, but then define $BH \cong \Lambda//G$ as $\frac{\Lambda \times \hat{E}G}{G}$, where $\hat{E}G := EG \times E(G/H_0)$. Note that $\hat{E}G$ is a weakly contractible space that admits a free action of $G$, which is the diagonal action induced from the action of $G$ on $EG$ and $E(G/H_0)$. Therefore it is legitimate to replace $EG$ with $\hat{E}G$ in the definition of $\Lambda//G$. Elements of $BH$ can now be expressed as classes $[\lambda, e, \tilde{e}]_G$, where $\lambda \in \Lambda, e \in EG, \tilde{e} \in E(G/H_0)$. $[x]_G$ indicates an orbit of $x$ under $G$ action.

The map $\tau \circ \pi$ takes $[\lambda, e, \tilde{e}]_G \rightarrow [\tilde{e}]_{G/H_0}$. The fiber of $[\tilde{e}]_{G/H_0}$ is therefore the set $\{[\lambda, e]_{H_0}|\lambda \in \Lambda, e \in EG\}$. Since $H_0$ acts trivially on $\Lambda$, this is equivalent to the set $\{(\lambda, [e]_{H_0})| \lambda \in \Lambda, e \in EG\}$, which is simply $\Lambda \times BH_0$.

\section{Homotopy-theoretic formulation of the compatibility relation}
\label{app:Homotopy}

Consider a space $\Lambda$ that is acted upon by a group $G$. We already described in Section \ref{sec:GC-Classif} above that if the $G$-action on the many-body Hilbert space is non-anomalous, families of gapped ground states over $\Lambda$ that is equivariant under the $G$-actions should be classified by homotopy classes of maps $\Lambda // G \to \Theta_d$, where $\Theta_d$ is the space of invertible ground states in $d$ dimensions.

Here we consider how to generalize this to the anomalous case. The idea is to exploit the fact that a system in $d$ dimensions with an anomalous action of $G$ can exist as the boundary of a $G$-SPT in one higher dimension with a non-anomalous $G$-action. In fact, in what follows, it will be more convenient to imagine that the system exists at the interface between two $G$-SPTs.

If the ground states at the interface remain invertible, then it should be possible to model the interface using the ``smooth state'' picture of Ref.~\cite{thorngren2018gauging}. Recall that an (invertible) smooth state on $\mathbb{R}^{d+1}$ is a continuous map
\begin{equation}
f : \mathbb{R}^{d+1} \to \Theta_{d+1},
\end{equation}
where $\Theta_{d+1}$ is the space of invertible states in $d+1$ spatial dimensions. Physically one can think of this as a state with a slow modulation as a function of space. If we want to describe a smooth state with a non-anomalous internal $G$ symmetry, then one would instead consider a map
\begin{equation}
    f : X \times BG \to \Theta_{d+1}.
\end{equation}
 However, in our case we want to consider a \emph{family} of smooth states over $\Lambda$ that is equivariant with respect to $G$. If the $G$-action on the many-body Hilbert space is non-anomalous, this amounts to considering maps
\begin{equation}
    f : (\Lambda \times EG)/G \times \mathbb{R}^{d+1} \to \Theta_{d+1}.
\end{equation}
 Now, we want to describe a planar interface in which the spatial variation happens only near the interface. Therefore, by an appropriate choice of coordinates, we can demand that $f(\cdot,\cdot,(x_1, \cdots, x_{d+1}))$ depends only on $x_{d+1}$ and is independent of $x_{d+1}$ for $x_{d+1} \leq 0$ or $x_{d+1} \geq 1$. Moreover, we want the $G$ symmetry to be broken only near the interface, which amounts to requiring that $f(\lambda, e, x_{d+1})$ is independent of $\lambda$ for $x_{d+1}= 0$ and $x_{d+1} = 1$.
All of this amounts to saying that we want to consider maps
\begin{equation}
\label{eq:interface_f}
f : (S\Lambda \times EG)/G \to \Theta_{d+1},
\end{equation}
where $S\Lambda = (\Lambda \times [0,1]) / (\Lambda \times \{0\}) / (\Lambda \times \{1 \})$ is the unreduced suspension of $\Lambda$, and inherits the $G$-action from $\Lambda$.

Note that there are two ``special'' points $s_0, s_1 \in S\Lambda$ corresponding to the image in the quotient space of $\Lambda \times \{0\}$ and $\Lambda \times \{1 \}$. From this we obtain two inclusion maps $EG \to S\Lambda \times EG$; since these are $G$-equivariant we obtain maps $\sigma_0, \sigma_1 : BG \to (S\Lambda \times EG)/G$. Thus, any map $f$ of the form \eqnref{eq:interface_f} defines two maps $f_0, f_1 : BG \to \Theta_{d+1}$. Physically, this defines the $G$-SPTs on the two sides of the interface.

    On the other hand, if $H_0$ is a subgroup of $G$ that acts trivially on $\Lambda$, then there is a natural map\footnote{This follows by a similar argument to Footnote 1 in Appendix A of Ref.~\cite{else2021goldstone}.} $S\Lambda \times BH_0 \to (S\Lambda \times EG)/G$. Composing this with $f$ gives us a map $\hat{f} : S\Lambda \times BH_0 \to \Theta_{d+1}$. Defining $\Theta_{d+1}^{H_0} = \mathrm{Map}[BH_0,\Theta_{d+1}]$, that is, the space of continuous maps from $BH_0$ to $\Theta_{d+1}$, then this is equivalent to a map $\hat{f} : S\Lambda \to \Theta_{d+1}^{H_0}$.
    As before, we can restrict to the points $s_0, s_1 \in S\Lambda$ to obtain elements $\omega_0, \omega_1 \in \Theta_{d+1}^{H_0}$. Because $s_0$ and $s_1$ are connected by a path in $S\Lambda$, it follows that $\omega_0$ and $\omega_1$ are continuously connected in $\Theta_{d+1}^{H_0}$, which physically corresponds to the statement that in order for the interface to remain gapped, it must be the case that the bulk states on either side of the interface are in the same $H_0$-SPT phase.

    Let us now assume that we have chosen the bulk states such that they are in the \emph{trivial} SPT phase with respect to $H_0$. What this means is that there must exist paths $\gamma_0, \gamma_1 : [0,1] \to \Theta_{d+1}^{H_0}$ that continuously connect $\omega_0$ and $\omega_1$ to $\omega_* \in \Theta_{d+1}^{H_0}$, a fixed \emph{constant} map from $BH_0 \to \Theta_d$. By combining these with the map $\hat{f}$, we can obtain a map $\hat{f}' : \Lambda \to \Omega \Theta_{d+1}^{H_0}$, where $\Omega \Theta_d$ is the based loop space of $\Theta_{d+1}^{H_0}$, i.e.\ the space of loops in $\Theta_{d+1}^{H_0}$ that begin and end at a fixed base point (in the case of $\Theta_{d+1}^{H_0}$, we choose $\omega_*$ as the base point). 

    Now we assume the space of invertible states in $d$ spatial dimensions satisfies the $\Omega$-spectrum property: $\Omega \Theta_{d+1}$ is (canonically) homotopy equivalent to $\Theta_d$. It is straightforward to show that this also implies that $\Omega \Theta^{H_0}_{d+1}$ is homotopy equivalent to $\Theta^{H_0}_d$. Therefore, from the map $\hat{f}'$ we obtain a map $\Lambda \to \Theta^{H_0}_d$, which defines an $H_0$-family over $\Lambda$.

    The one problem is that the resulting $H_0$-family can depend on the choice of paths $\gamma_0, \gamma_1$, which are not unique. This is a reflection of the torsoriality issue that we described in Section \ref{subsec:general_invertible}. However, one can show that if we quotient out by the constant $H_0$-families, as described there, the resulting pump invariant does not depend on the choice of $\gamma_0$ and $\gamma_1$.

\

\bibliography{refs_families}

\end{document}